\documentclass[useAMS,usenatbib]{mn2e}

\title[Tidally Induced  Spiral Structure]
{Slowly Breaking Waves: The Longevity of Tidally Induced Spiral Structure}
\author[C. Struck, C. L. Dobbs, and J.-S. Hwang] 
{Curtis Struck,\thanks{E-mail: curt@iastate.edu (CS);
cdobbs@mpe.mpg.de (CLD); jshwang@iastate.edu (J-SH)}$^1$
Clare L. Dobbs,$^{2,3,4}$
Jeong-Sun Hwang$^{1,5}$ \\
$^1$ Dept. of Physics and Astronomy, Iowa State Univ., Ames, IA, 50014 USA\\
$^2$ School of Physics, University of Exeter, 
Stocker Road, Exeter, EX4 4QL, UK \\
$^3$ Max-Planck-Institut f\"ur extraterrestrische Physik, Giessenbachstra\ss{}e, D-85748 Garching, Germany \\
$^4$ Universitats-Sternwarte M\"unchen, Scheinerstra\ss{}e 1, D-81679 M\"unchen, Germany\\
$^5$ School of Physics, Korea Inst. for Advanced Study, Seoul 130-722, Republic of Korea}
\usepackage{amssymb}
\usepackage{amsmath}
\usepackage{graphicx}
\usepackage{epsfig}
\usepackage{multirow} 
\def\aap{{ A\&A}}

\def\aj{{AJ}}

\def\apj{{ApJ}}
\def\apjl{{ApJL}}

\def\mnras{{MNRAS}}

\def\apjs{{ApJS}}

\begin{document}
\date{\today}

\pagerange{\pageref{firstpage}--\pageref{lastpage}} \pubyear{0000}

\maketitle

\label{firstpage}
\begin{abstract}
We have discovered long-lived waves in two sets of numerical models of fast (marginally bound or unbound) flyby galaxy collisions, carried out independently with two different codes.  In neither simulation set are the spirals the result of a collision-induced bar formation. Although there is variation in the appearance of the waves with time, they do not disappear and reform recurrently, as seen in other cases described in the literature. We also present an analytic theory that can account for the wave structure, not as propagating transients, nor as a fixed pattern propagating through the disc. While these waves propagate through the disc, they are maintained by the coherent oscillations initiated by the impulsive disturbance. Specifically, the analytic theory suggests that they are caustic waves in ensembles of stars pursuing correlated epicyclic orbits after the disturbance. This theory is an extension of that developed by Struck and collaborators for colliding ring galaxies.

The models suggest that this type of wave may persist for a couple of Gyr., and galaxy interactions occur on comparable timescales, so waves produced by the mechanism may be well represented in observed spirals. In particular, this mechanism can account for the tightly wound, and presumably long-lived spirals, seen in some nearby early-type galaxies. These spirals are also likely to be common in groups and clusters, where fast encounters between galaxies occur relatively frequently. However, as the spirals become tightly wound, and evolve to modest amplitudes, they may be difficult to resolve unless they are nearby. Nonetheless, the effect may be one of several processes that result from galaxy harassment, and via wave-enhanced star formation contribute to the Butcher-Oemler effect.  
\end{abstract}


\section{Introduction}
The nature of spiral structure in galaxies is a long-standing problem
in astrophysics. The latest simulations now show that gravitational
instabilities in the stars lead to flocculent and multi-armed spirals
which persist for many Gyr (\citealt{Fujii2010}, \citealt{Oh2008}). However 
the mechanism which produces and maintains two-armed grand design galaxies is still ambiguous. 
We investigate in this paper whether interactions can induce a persistent
 m=2 pattern in spiral galaxies.


Grand design galaxies, which exhibit a symmetric two-armed spiral
structure, represent a significant fraction of spiral galaxies, perhaps
50\% \citep{Elmegreen1983}. The challenge of producing such a spiral
galaxy faces two major obstacles, firstly inducing the $m=2$ spiral
structure, and secondly maintaining it. The most obvious means of
inducing an m=2 spiral is via an interaction. Many galaxies,
e.g. M51 and M81, are or have recently been involved in an
interaction with one or more companions, which explains their current spiral structure. 
Another possibility is that $m=2$ spiral structure may be driven by a bar
\citep{Huntley1978,Roberts1979,Schempp1982,RG2007,Buta2009}. In early observational surveys \citet[also \citealt{Elmegreen1983}]{Kormendy1979} confirmed the frequent association between spirals and companions or bars. However it is not generally believed that all $m=2$ spirals are the
result of interactions or bars. A common argument against interactions
as the sole origin for non-barred $m=2$ spirals is the presence of isolated grand
design galaxies, e.g. M74. Nevertheless it is often difficult to tell
how isolated these galaxies really are. For example M74 is circumposed
by two massive HI clouds, which could themselves instigate the spiral
structure, or suggest a recent interaction
\citep{Kamphuis1992,Bottema2003}. Furthermore it is unclear how long
spiral structure persists after an interaction, and how far a
perturbing galaxy may have travelled. Observations of nearby galaxies
in fact suggest that their spiral morphology has a tidal origin, since
the spiral arms have radially decreasing rather than constant pattern speeds \citep{Meidt2009}.

An alternative scenario is that the spiral arms are due to
quasi-stationary spiral density waves (see review of \citealt{se10}). Even then, bars or
interactions are often presumed to invoke these spiral waves, and it is not
readily possible (e.g. with numerical simulations) to demonstrate that quasi-stationary spiral density
waves can be induced in an isolated galaxy. \citet{Toomre1969} showed
that for the quasi-steady spiral structure hypothesised by
\citet{Lin1964}, spiral density waves are quickly damped. Thus some
mechanism is required to maintain the amplitude of the spiral density
waves, one suggestion being swing amplification
\citep{Toomre1981,Mark1976}. The waves also need to be reflected at
the centre of the galactic disc, requiring that there is no inner
Lindblad resonance which would absorb such waves. These conditions
necessitate a fairly high disc to halo ratio, since a massive halo
suppresses swing amplification. However such galaxies are not stable
and naturally lead to the formation of a bar. 
The simulations that most closely reproduce an isolated $m=2$ galaxy
are those of \citet{Sellwood1985}, \citet{Thomasson1990}, and \citet{zh98}, who manage to
prevent bar formation with the presence of  a bulge. In these
calculations an $2 \le m \le 4$ spiral galaxy develops, where $m$ typically varies with time.

To determine whether grand design spiral structure in non-barred galaxies can be explained entirely by
interactions requires estimating the frequency of interactions, and
the duration of grand design structure. Slow collisions result in a prompt merger; fast collisions between galaxies have been little studied, so the question of wave persistence remains open. Perturbations of short duration have much less effect than those between comparable galaxies in longer encounters, e.g., see the reviews of \citet{st99} and \citet{st06}. Exceptions to this generalization include direct collisions, where proximity offsets brevity, and cases where multiple fast, weak disturbances in a cluster environment might have a cumulative effect, the so-called harassment effect \citep{mo96}. Certainly in fast collisions with unbound or marginally bound partners there has been little motivation to look for long-term effects.

Here we investigate the question of the duration of induced spirals by performing two sets of numerical simulations of interactions. The first set consists of relatively simple simulations using rigid halo potentials and a local self-gravity within the galaxy discs. This facilitates study of the long-term evolution of waves in the discs. Our second set of models consists of fully self-consistent, high-resolution N-body hydrodynamic models. In a recent paper \citet{Dobbs2010} modelled the interaction of M51 and NGC 5195 to reproduce the current spiral structure of M51. The adopted orbit was bound, so the grand design structure only lasts around 200 Myr before the two galaxies merge. In the second simulation set, we use this code to perform similar calculations to those of M51, but give the companion a higher velocity initially, so the orbit is unbound. We also vary the ratio of the companion to the primary galaxy. 

Specifically, the models of Set 1, with their fixed potentials and modest dynamical heating, produce very well defined arms, and allow longer timescales to be studied. The models of Set 2, allow gas dynamics to be studied in more detail, and are generally more self-consistent. Close orbits with a lower mass companion are considered in Set 2. In Set 1 the orbits are more distant, but with a higher mass perturber. Thus, a fairly wide range of conditions and approximations are tested with the two codes. 

\citet{Oh2008} also published similar simulations on flyby galaxy interactions, and analysed the properties of the resulting tidal structures. We differ from their calculations by including gas, and the calculations in Set 2 use a live halo, and are three-dimensional (these calculations also do not assume that one galaxy is fixed).  In the current paper, we also focus more on the lifetimes of spirals induced in the models, and explaining them in a theoretical and observational context.

In the penultimate section we present an analytic model, which allows us to interpret both models and qualitative features of the observations. For the last forty years or more, most analytic work on spiral structure in galaxy discs has been based on the linear perturbation analysis of small amplitude waves of an assumed form (see \citealt{bt}). The theory described below does not assume fixed forms for the wave patterns or propagating wave packets as in the classical theory. It also does not assume ongoing driving by bars or external forces beyond the initial impulsive disturbance. Rather it is based on the propagation characteristics of nonlinear, caustic waves (or gas shocks) induced by the tidal disturbance. The theory of such waves is well developed for symmetric ring galaxies \citep{ap96, struck2010}, and has been explored semi-analytically in asymmetric waves (e.g., \citealt{su87, sm90, ge94}). 

The comparison between analytic and numerical models is good, and the insights derived from that comparison can help us understand several observational conundrums. These include wave longevity and others that will be described in the final sections. One that is worth mentioning here is the existence of tightly wound spirals. These objects are reasonably well represented in catalogs of nearby galaxies, so they are evidently not too rare. NGC 488 is a prototypical example from the Hubble Atlas \citep{sa61}. The existence of tight spirals is not usually considered a puzzle, but on examination their place in density wave theory seems ambiguous. Naively,  we may simply be seeing the predicted windup of the spiral pattern. However, the morphology of these galaxies does not look like the windup of a traveling wave around a Lindblad resonance. Alternately, it may be the global windup of a quasi-steady pattern. However, recent work suggests that gaseous dissipation may be needed for the persistence of undriven spirals \citep{ch08}. Gas-poor, early-type galaxies would seem an unlikely site for wrapped waves if that is true. Of course, the potential in galaxies with large bulges may favor tight winding from the outset,  as expected with a declining rotation curve (see e.g., \citealt{bt}). The length, coherence, and large radial range of the observed cases give at least the appearance of longevity.  These characteristics can be accounted for in the theory described below. 


\section{Numerical Models: Set 1}
\subsection{Description of the Numerical Code}

The basic code used to produce our first set of simulations was a slightly modified version of that of \citet{st97}. This is an SPH gas dynamics  code, though the hydrodynamics does not play a large role in the present discussion. A simple leapfrog integrator is used to advance the stellar orbits. The gravitational potential of each galaxy is modeled with fixed, rigid halo potentials of the softened power-law form described in \citet{Sm2008}. Specifically, the potential of the model disc is such that its rotation curve is linearly rising in the core, and flat at radii much larger than the softening length of one unit. The potential of the companion is of the same form, except moderately declining at large radii.

Forces are computed on a fixed grid for computational simplicity, and local self-gravity in the disc is computed between particles in adjacent grid cells. Because of the disc shear it is negligible on much larger scales, when large disturbances break the symmetry of the disc. The grid size is 0.05 code units. We adopt a physical length scale of 2 kpc and time scale of 333 Myr. Then the outer orbital period of about 0.8 units equals 270 Myr. A total of 38,100 particles were used to model the primary star disc, whose initial size was about 9.0 kpc in the adopted units. The flat rotation curve velocity of the disc was 220 km s$^{-1}$, derived from a scale mass of $2.2 \times 10^{10}$ M$_\odot$ within a radius of 2.0 kpc. The companion consisted of a halo potential only.

The companion's mass was taken to be about twice that of the target galaxy. The initial position of the companion was about 120 kpc south of the primary with initial velocity components of 150 km s$^{-1}$ west (toward positive x values) and 600 km s$^{-1}$ northward. The relatively high value of the companion mass was set to achieve a significant perturbation at the relatively large distance of closest approach, which was about 2.6 times the radius of the initial disc, and the high flyby velocity. The companion orbit was only perturbed by an angle of about 20$^\circ$ in the encounter. As will be seen below, the Set 1 models show less smoothing than the full N-body models of Set 2. The random velocity components were initialized to low values in the former to  allow us to discern their long-time development more clearly; the smoothing of the self-consistent Set 2 models is more realistic.

\subsection{Set 1 Model Results}

 Figure 1 shows the evolution of waves in the model disc stars, with time measured in code units from the time of closest approach between the two galaxies. The model results shown in Figure 1 provided several surprises.

\begin{figure*}
\centerline{
\includegraphics[scale=0.27]{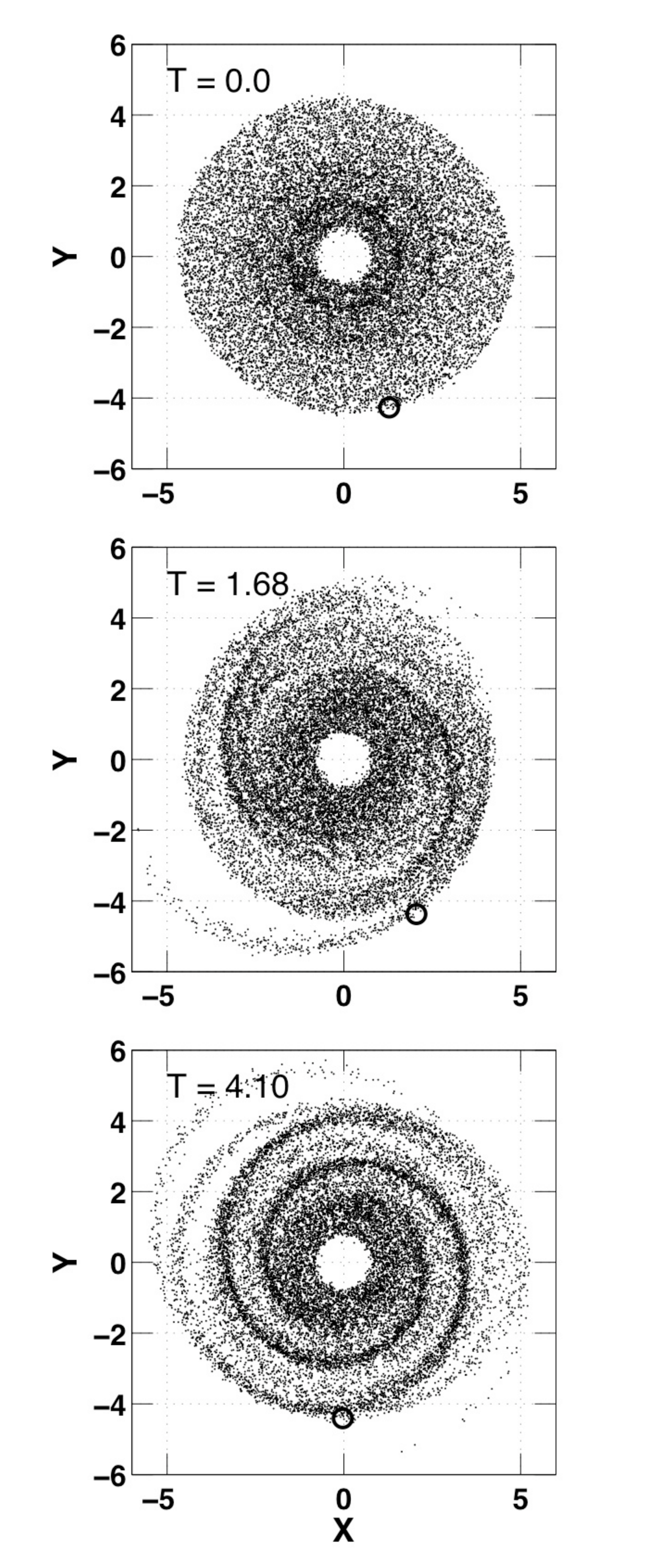}
\includegraphics[scale=0.27]{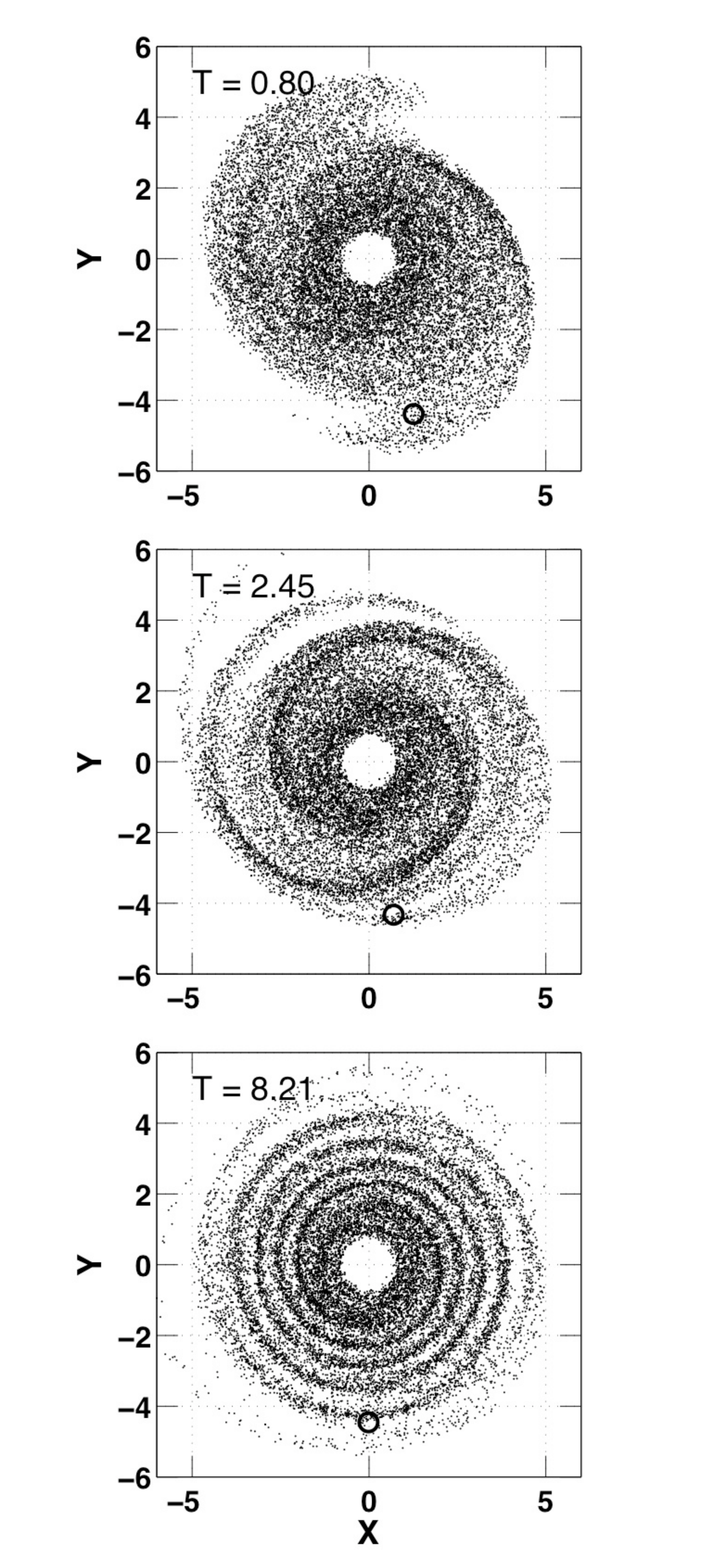}}
\caption{The numerical model at 6 times. The zero time in the first panel corresponds to the closest approach of the companion. Code time units are used as described in the text. The large circle marks the location of a particular outer disc particle. The second panel shows the disc about one outer disc rotation time after the encounter (270 Myr in the adopted scaling). The subsequent panels show the disc after about two, three, five and ten outer disc rotation periods.}
\end{figure*}  
  
The early response of the disc is very mild, as expected. The first surprise is that the spirals do not dissipate in a few (outer disc) orbital times. The rapid disappearance of spiral waves by wind-up or dissipation has been a problem noted in many simulation studies. Over the years there have been various suggestions about how this fate might be avoided. These include the stabilizing effects of gaseous dissipation, e.g., \citet[and references therein]{ch08}, and the offset between the extrema of the density and gravitational potential in a wave \citep{zh98}. However, the figures from the simulations in these works suggest that, in fact, the ameliorating effects of these processes are moderate at best.  On the other hand, \citet{se10} has suggested that spirals are transient, but that they continuously regenerate via a recurring cycle of instability. Even this is not quite the same as the longevity of the tidally induced arms in our simulations.

In Figure 1 the density contrasts across the spiral density waves grow stronger for a time, and the waves do not wind up at a particular (e.g., resonant) radius. This growth is not as rapid as in the case of swing amplification in discs with low values of the \citet{Toomre1981} $Q$ parameter. (The dimensionless Toomre parameter $Q = {\kappa}{\sigma}/(3.36G{\Sigma})$ for stars, with surface density $\Sigma$, epicyclic frequency $\kappa$, and velocity dispersion $\sigma$, implies local gravitational instability at values less than about 1.0 (see \citealt{bt}).) The waves are greatly stretched due to the shear in the disc. This is expected from classical density wave theory \citep[and references therein]{bt}, and as seen in many simulation studies, which generally have stronger initial wave amplitudes. The second surprise is that the after this initial wave steepening, or Òbreaking,Ó the waves continue to persist for a much longer time (more than 10 outer orbital periods in the model of Fig. 1) while winding ever tighter. A number of checks were made to eliminate the possibility of numerical errors in the code or graphical errors. We will see in Section 4 that the results can be accounted for with analytical calculations. 

The goal of the Set 1 models was to provide a clear picture of the evolution of tidally induced waves. Feedback effects such as energy and momentum inputs from star-forming regions (also included in the code), or scattering by self-gravitating clumps tend to smooth out the disc and obscure waves. Local self-gravity is included in the code, but to minimize these effects particle masses were reduced to obtain a high value of the Toomre Q parameter. Specifically, these models had $Q \simeq 10$. Although quite high, such a $Q$ value is appropriate for the outer parts of early-type galaxy discs (e.g., \citealt{Kennicutt1989}). However, more realistic model discs with characteristics like those in later type galaxies are considered in the next section.

\section{Numerical Models: Set 2}
\subsection{Description of the Models}
The calculations described in this section use an SPH code, originally developed
by \citet{Benz1990}, but with substantial modifications, such as
individual particle timesteps, grad(h) implementation, magnetic fields
and sink particles. The code is predominantly used for simulating gas
dynamics, but we have modified the code to include stellar (or dark
matter) particles as well \citep{Dobbs2010}.  
In all calculations, we include particles for a gaseous and stellar
disc, bulge and halo. The setup of the simulations is similar to that
described in \citet{Dobbs2010}, using the NEMO package \citep{Teuben1995} to obtain the
initial particle positions and velocities. We used 1 million particles
for the disc, which comprises 900,000 gas particles and 100,000
stellar particles, 100,000 particles for the halo and 20,000 for the
bulge. We thus have a slightly better resolved halo compared to
\citet{Dobbs2010}.

Otherwise the main difference from \citet{Dobbs2010} is that the
initial positions and velocities of the perturbing galaxy are altered
to produce an unbound orbit. The main galaxy is the same as the
previously modelled M51, so has a disc mass of $5.9 \times 10^{10}$
M$_{\odot}$, and halo mass of $1.45 \times 10^{11}$ M$_{\odot}$, and a
small bulge of mass $5.25 \times 10^{9}$ M$_{\odot}$. This results in
a rotation curve with a maximum velocity of 275 km s$^{-1}$. 

The disc includes stars and gas, but we adopt a very low gas fraction, 1\% to avoid
gravitational collapse in the gas. Even so, we find that at later
times there is significant accretion into the inner parts of the disc,
requiring the insertion of sink particles, with the same criteria as \citet{Dobbs2010}.
The gas is isothermal, with a temperature of $10^4$ K, and is
initially distributed with the stars, though the gas quickly settles
to an equilibrium in the $z$ plane \citep{Dobbs2010}.

We describe the galaxy in terms of a Cartesian
coordinate system in which the $xy$ plane is equivalent to the plane
of the sky, and the $z$ direction lies towards us along the line of
sight. Similarly to our calculation of M51, we perform two rotations,
of $20^o$ about the $y-$axis, and $10^o$ about the $z-$axis.
Essentially we are aiming to establish how long spiral structure 
lasts in a system comparable to M51 and NGC 5195, but where 
the evolution is not terminated by a merger. Hence the 
variable values in this model are similar to the previous calculations, and so, resemble a realisitic physical system.  

The companion is modelled as a point mass. We perform
calculations with different masses for the companion galaxy, where the
ratio of mass of the companion to that of the main galaxy
is 0.01, 0.1 and 0.3, so unlike the Set 1 calculations, the perturber always has a smaller mass than the companion. The last case, 0.3, is the same ratio as employed
in the previous calculations designed to model M51. We provide the
initial positions and velocities of the two galaxies in
Table~1. Compared to \citet{Dobbs2010}, the companion galaxy now
starts further away, and with double the velocity in each
direction. We also show the orbits of the two galaxies in Figure~2, for
the 3 different mass cases. The orbit of the companion, which has a
much higher velocity, changes very little for the different cases, the
highest ratio companion  has a slightly reduced velocity. 
The orbit of the main galaxy is however much more perturbed for the highest ratio companion.

\begin{table}
\centering
\begin{tabular}{r|c|c|c}
 \hline 
& & Modelled galaxy & Companion \\
\hline
Initial & $x$ & 4.91 & -22 \\ 
position & $y$ & 1.89 & -11.55 \\ 
(kpc) & $z$ & 0.95 & -8.30 \\ 
\hline
Initial & $v_x$ & -1.46 & 11.0 \\ 
velocity & $v_y$ & 0.68 & -4.9 \\ 
(km s$^{-1}$) & $v_z$ & -3.26 & 31.0 \\ 
\hline
\end{tabular}
\caption{The initial positions and velocities are listed for the main galaxy, and the companion galaxy for the Set 2 simulations. The orbit is similar to that for the previous calculations of M51 \citep{Dobbs2010} except the velocity of the companion is double in each direction, so the orbit is unbound.}
\label{orbit}
\end{table}

\begin{figure}
\centerline{
\includegraphics[scale=0.35]{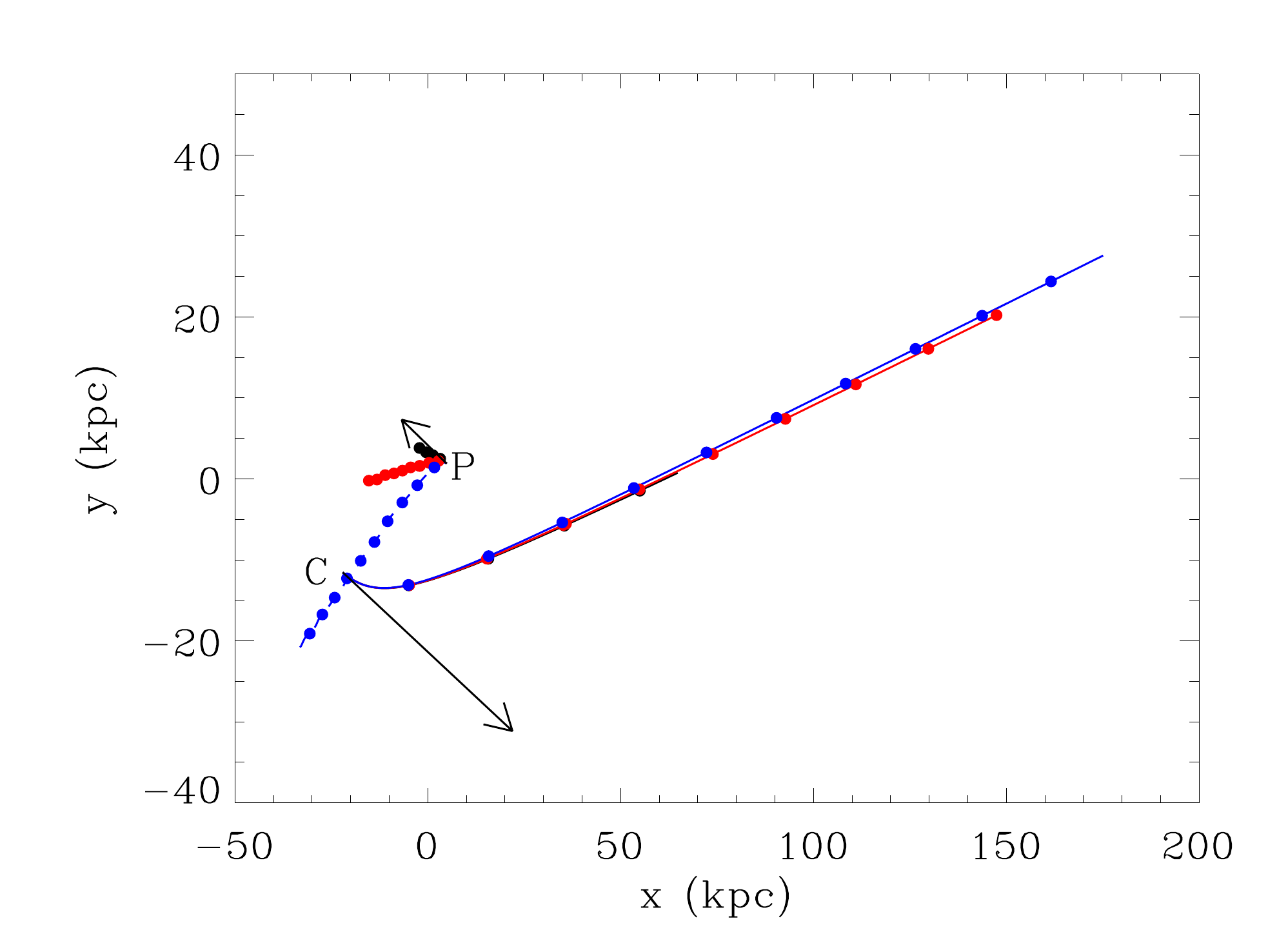}}
\centerline{
\includegraphics[scale=0.35]{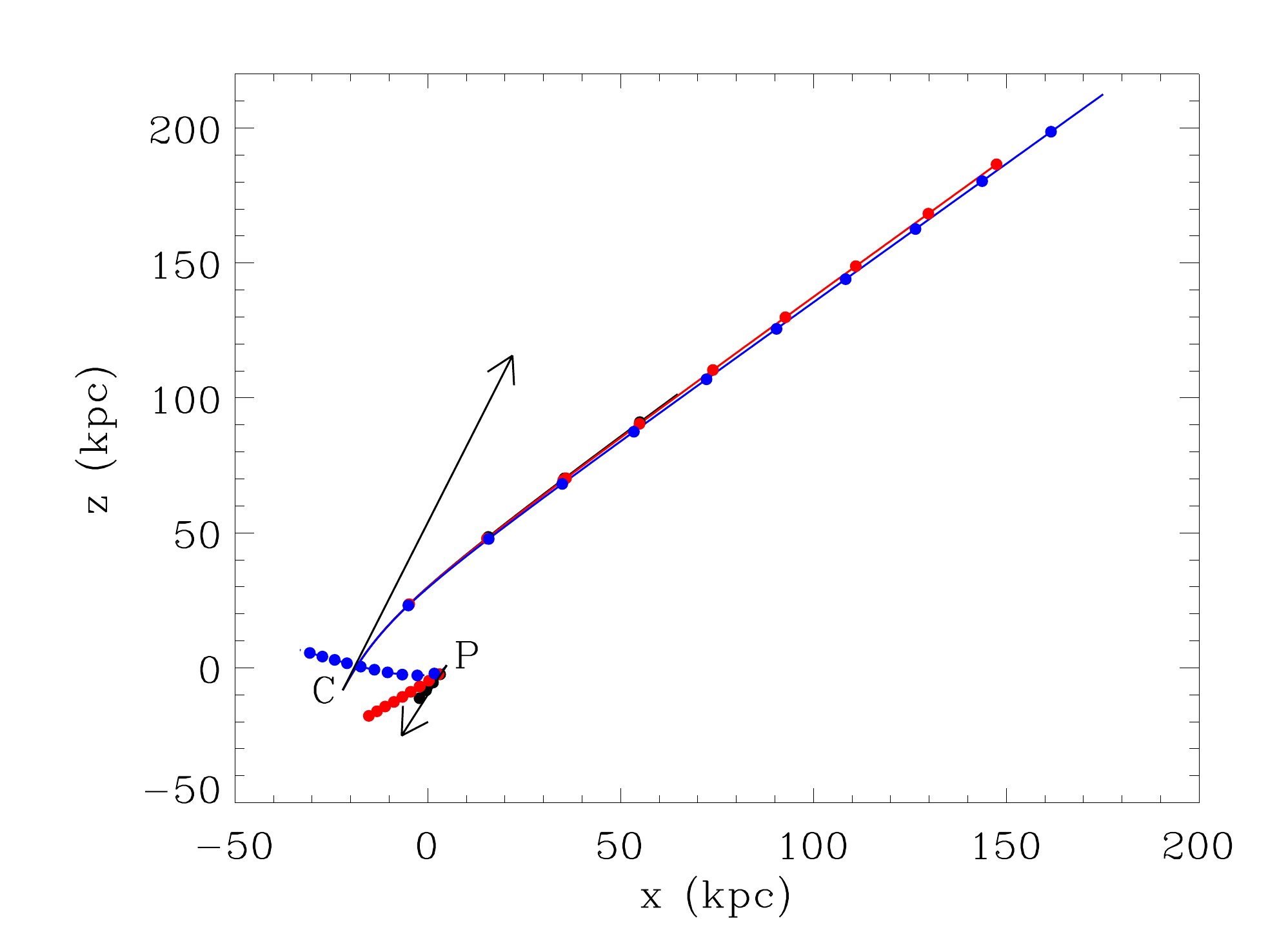}}
\caption{The orbits of the main galaxy and companion are shown where
  the ratio of the companion is 0.01 (black), 0.1 (red) and 0.3
  (blue). The orbits are shown in the $xy$ plane (i.e. face on) and
  $xz$ plane (lower). The `C' and `P' indicate the starting locations
  of the companion and primary galaxies, whilst the dots along the lines indicate time intervals of 100 Myr. The arrows represent the initial velocity vectors of the galaxies, but in order to fit in the plot, the length of the arrows for the companion has been halved relative to those of the main galaxy.}
\end{figure} 

\subsection{Results: Isolated case}
Before showing the main results for the interacting galaxies, we first show a case where the galaxy is isolated. As shown in Figure~3, the galaxy exhibits multiple long spiral arms. The galaxy does not form a bar, or an $m=2$ spiral due to a sufficiently massive halo, at least for 500 Myr (though a massive halo does not always prevent a bar forming, see \citet{Athanassoula2002}). The value of the Toomre Q factor of the stars is $1 \le Q \le 2$. Although we cannot confirm the longer term evolution, the 500 Myr exceeds the timeframe for the generation of two-armed spiral structure in the interacting cases.

\begin{figure}
\centerline{
\includegraphics[scale=0.25]{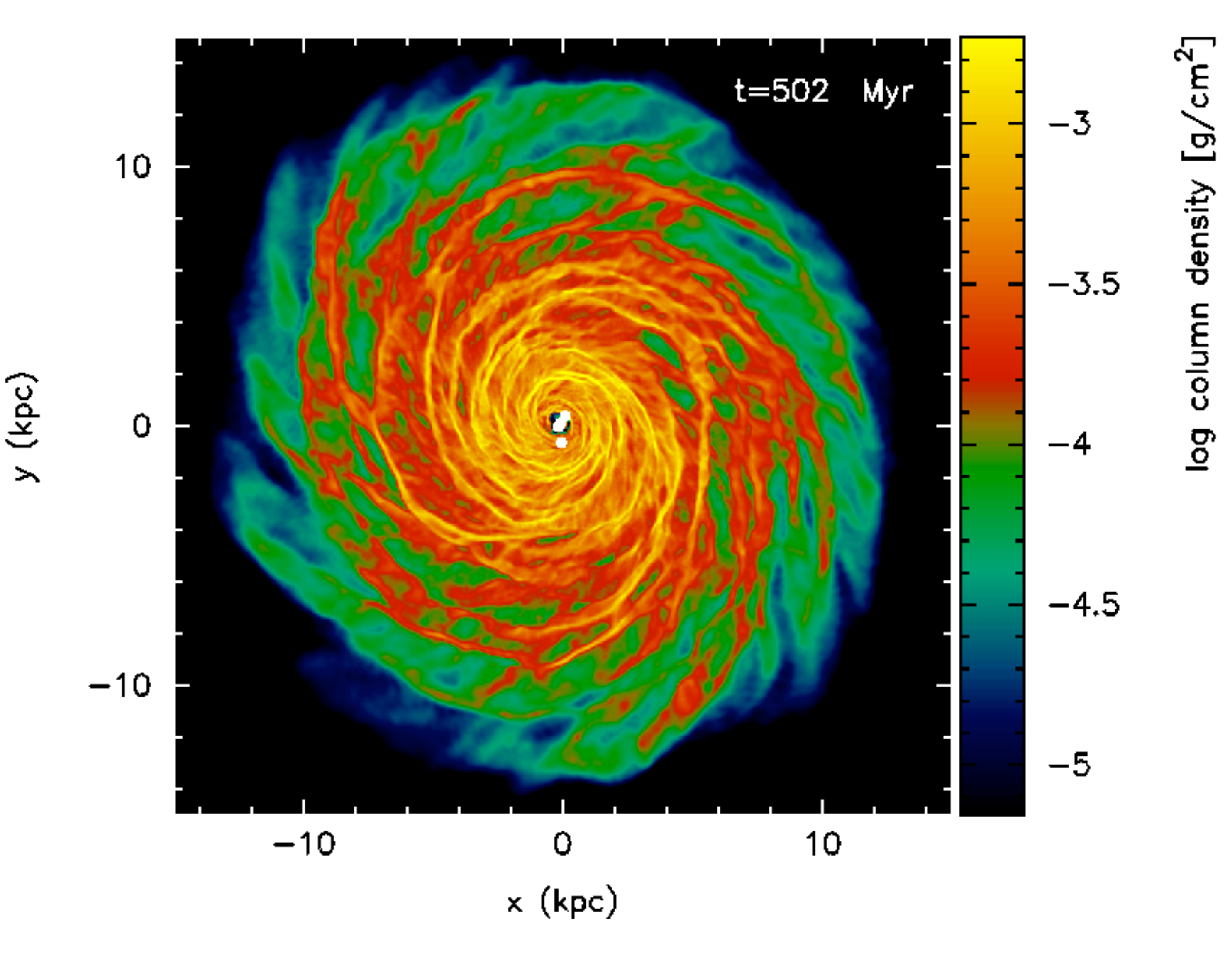}}
\caption{The column density of the gas for the isolated case is shown after 500 Myr. There is no evidence of an $m=2$ perturbation in the disc. In this, and the other Set 2 model figures, the white dots represent sink particles.}
\end{figure}  
    
\subsection{Results: Evolution of interacting galaxies}
\begin{figure}
\centerline{
\includegraphics[bb=150 50 600 660, scale=0.2]{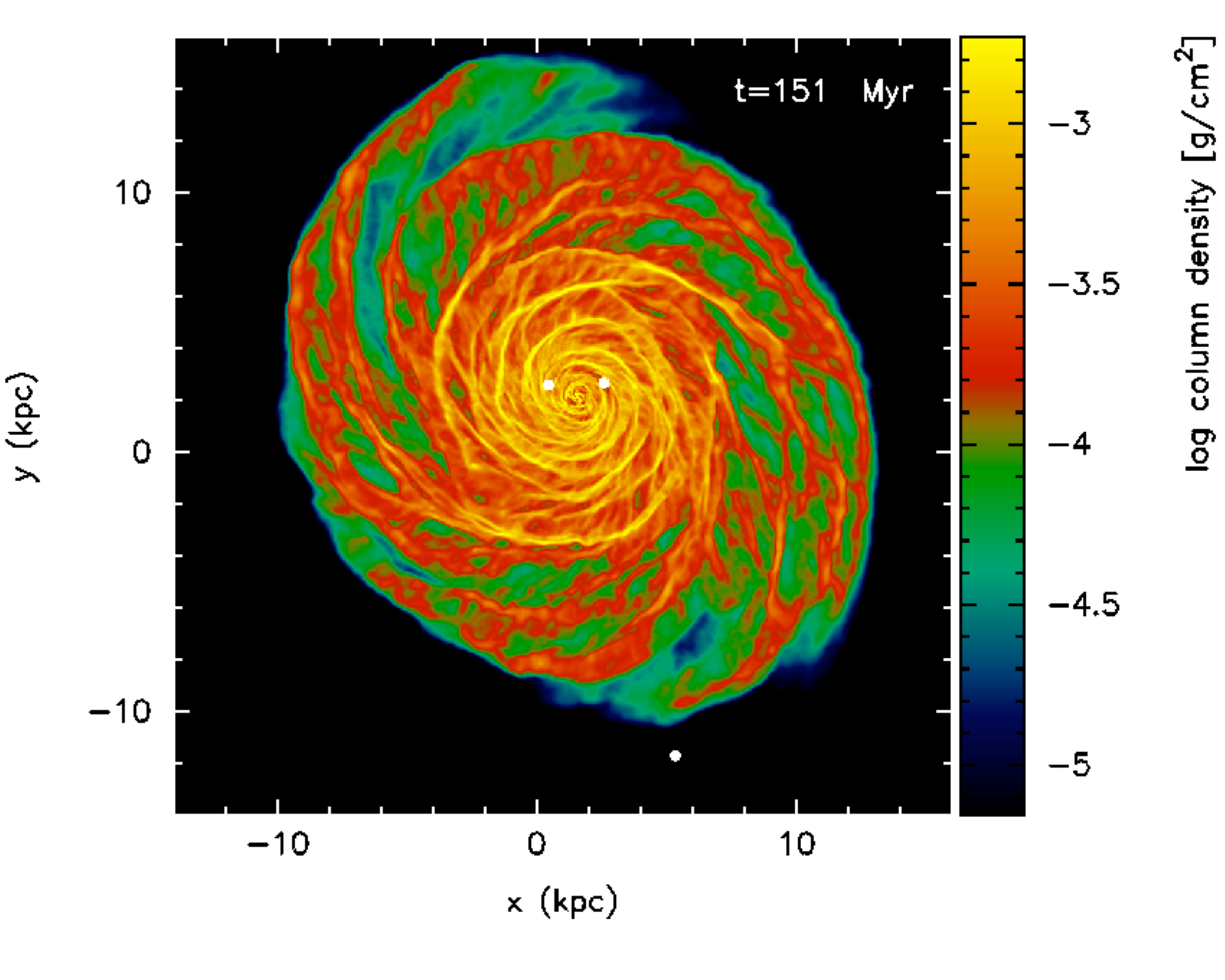}}
\centerline{
\includegraphics[bb=150 50 600 680, scale=0.2]{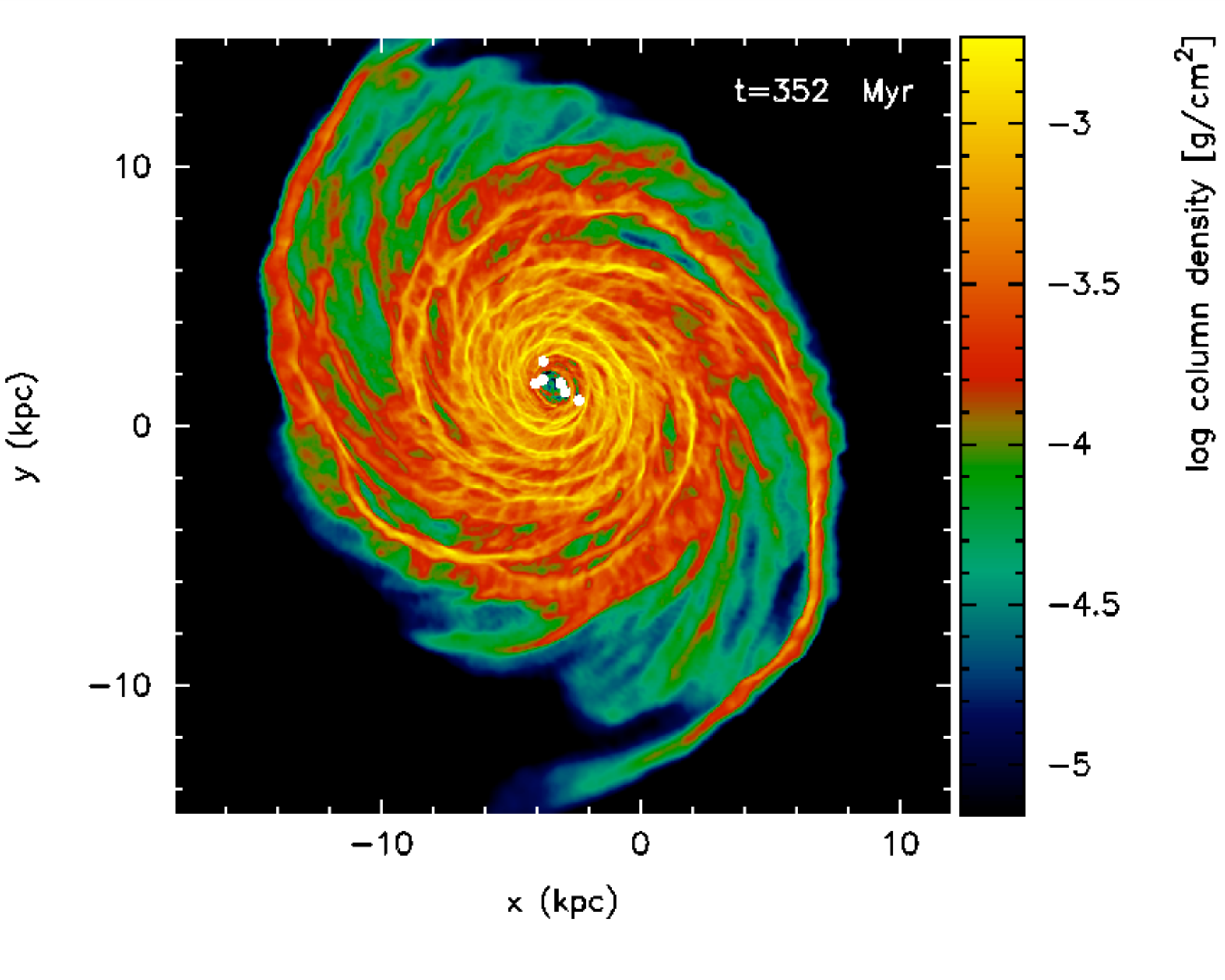}}
\centerline{
\includegraphics[bb=150 50 600 680, scale=0.2]{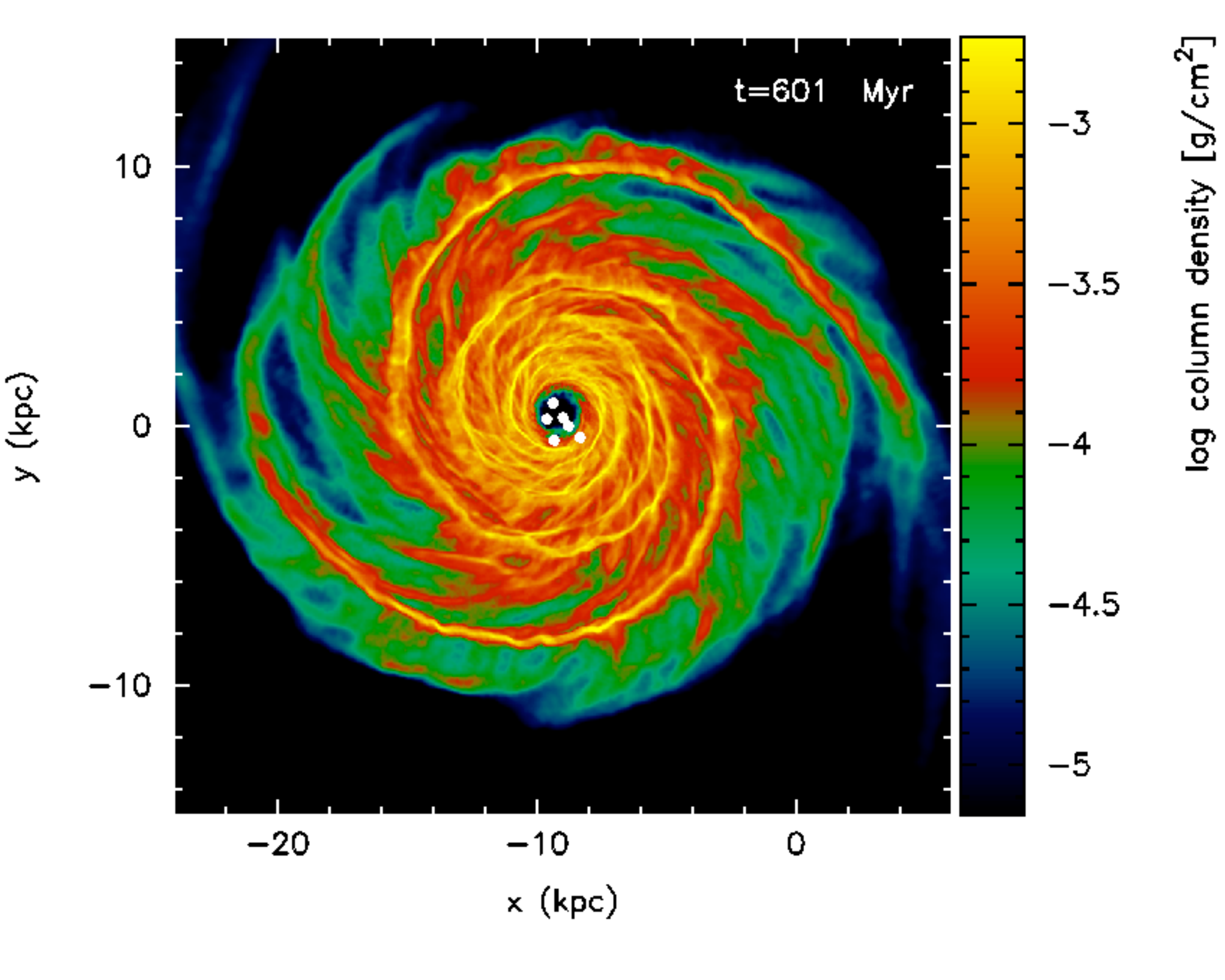}}
\centerline{
\includegraphics[bb=150 50 600 680, scale=0.2]{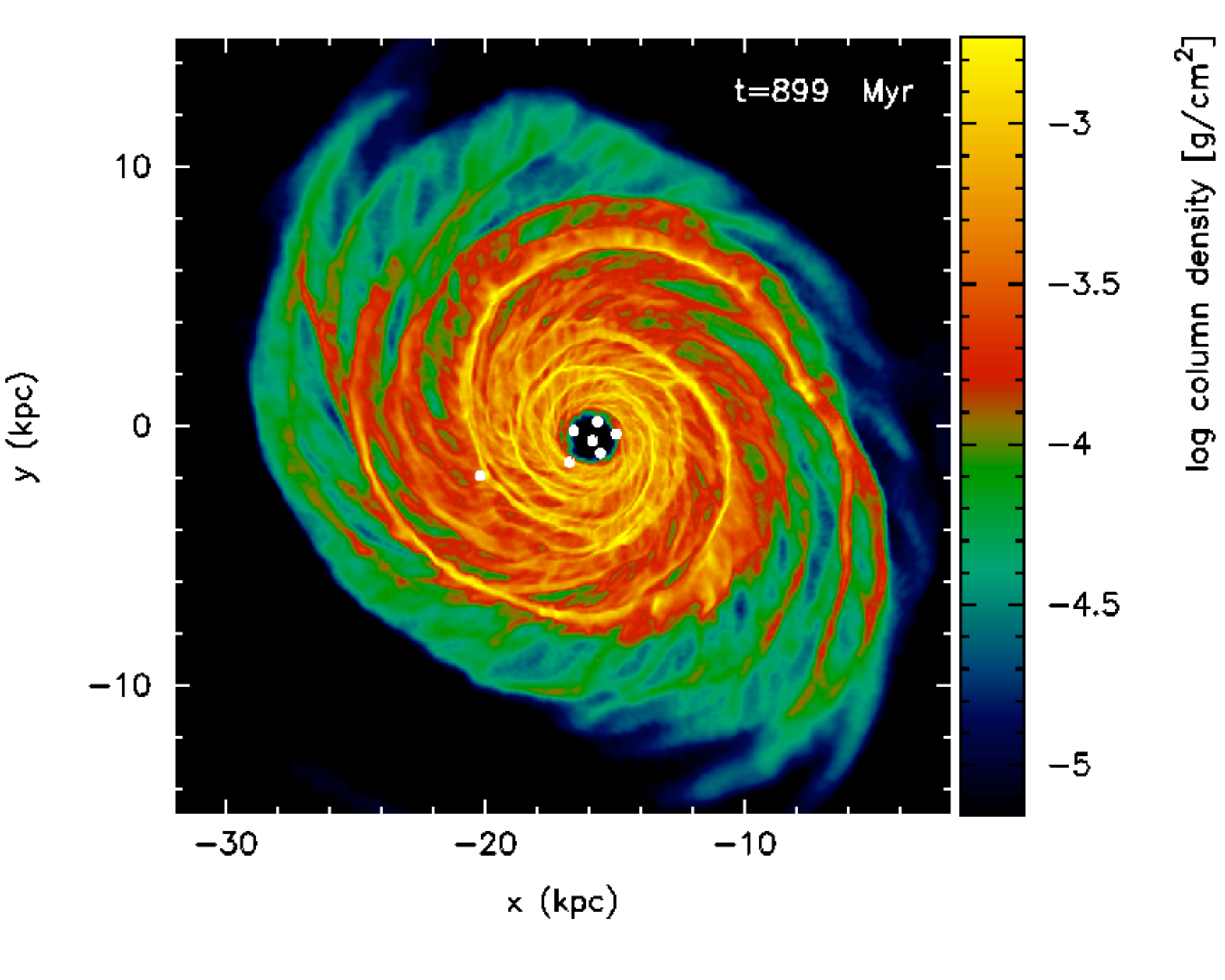}}
\caption{These panels show the evolution of the gas density for the
  0.1 ratio companion, at times of 151, 352, 601 and 899 Myr. There is
  a clear $m=2$ perturbation, although this is not induced until some
  time after the companion has passed. The perturbation lasts for
  at least 500 Myr, over which the arms of the galaxy gradually
  wind up.}
\end{figure} 
  
\begin{figure}
\centerline{
\includegraphics[bb=150 50 600 680, scale=0.2]{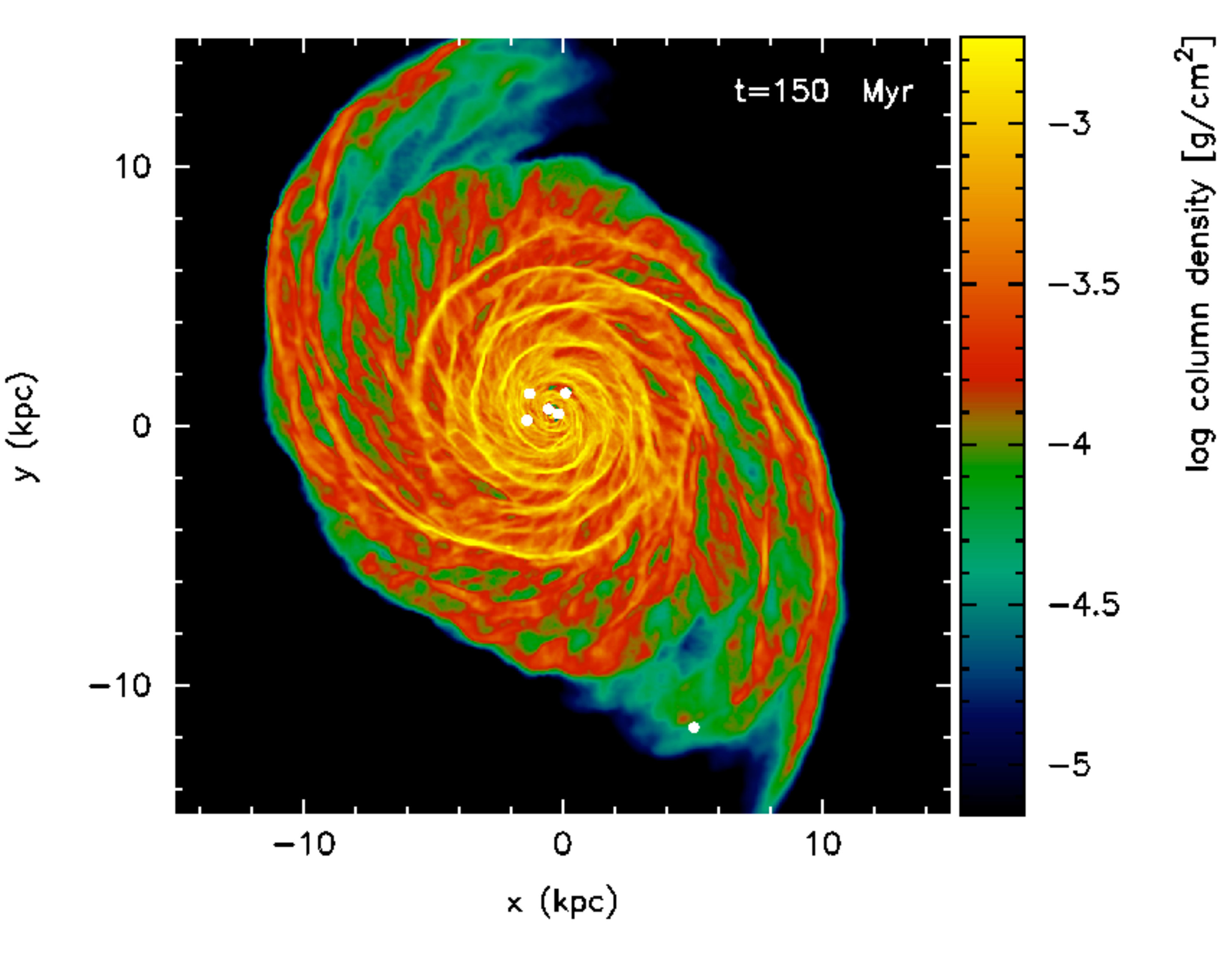}}
\centerline{
\includegraphics[bb=150 50 600 680, scale=0.2]{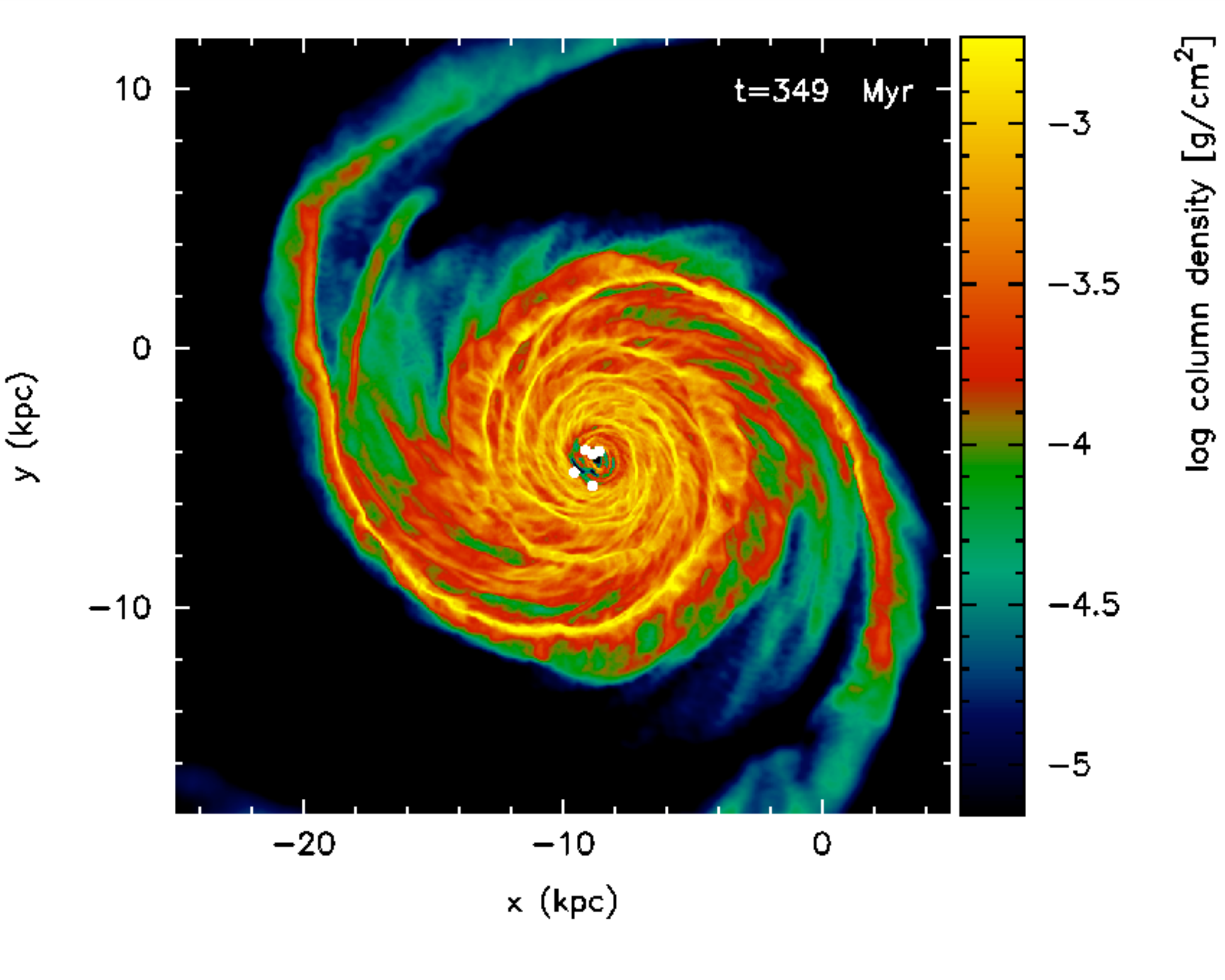}}
\centerline{
\includegraphics[bb=150 50 600 680, scale=0.2]{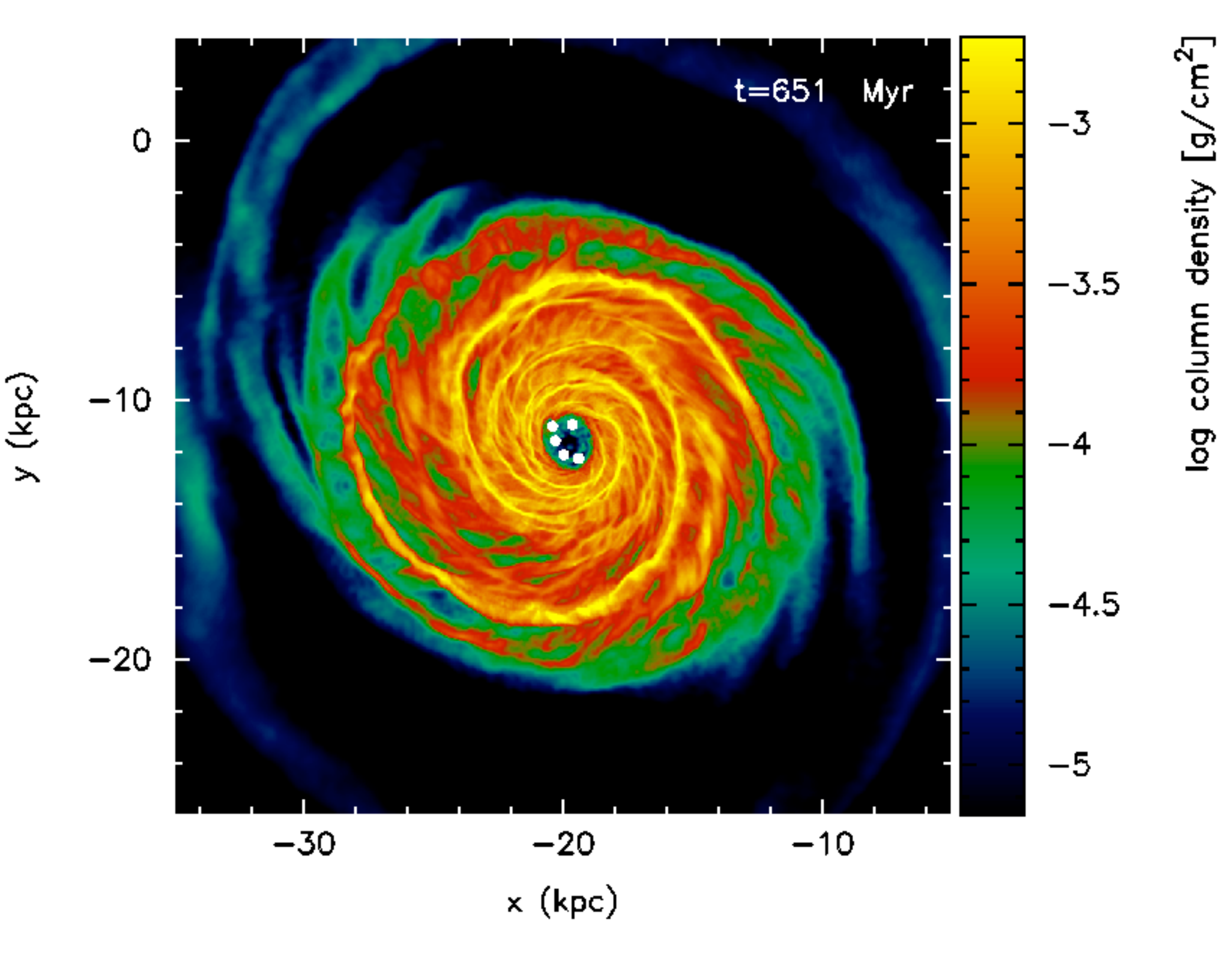}}
\centerline{
\includegraphics[bb=150 50 600 680, scale=0.2]{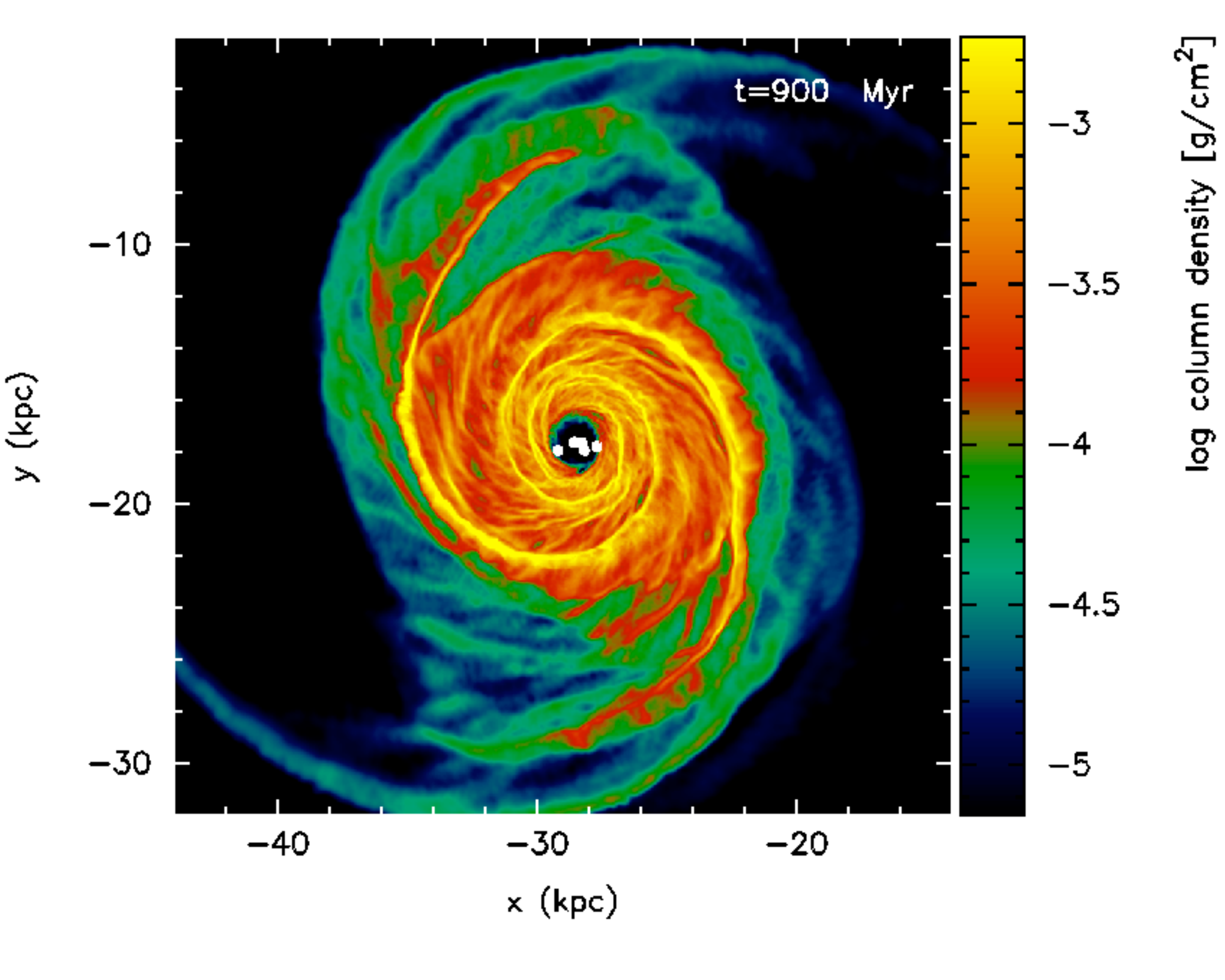}}
\caption{The evolution is shown for the 0.3 mass ratio companion. There is evidently a strong $m=2$ perturbation,  which again lasts for hundreds of Myr.}
\end{figure}
 
\begin{figure}
\centerline{
\includegraphics[scale=0.25]{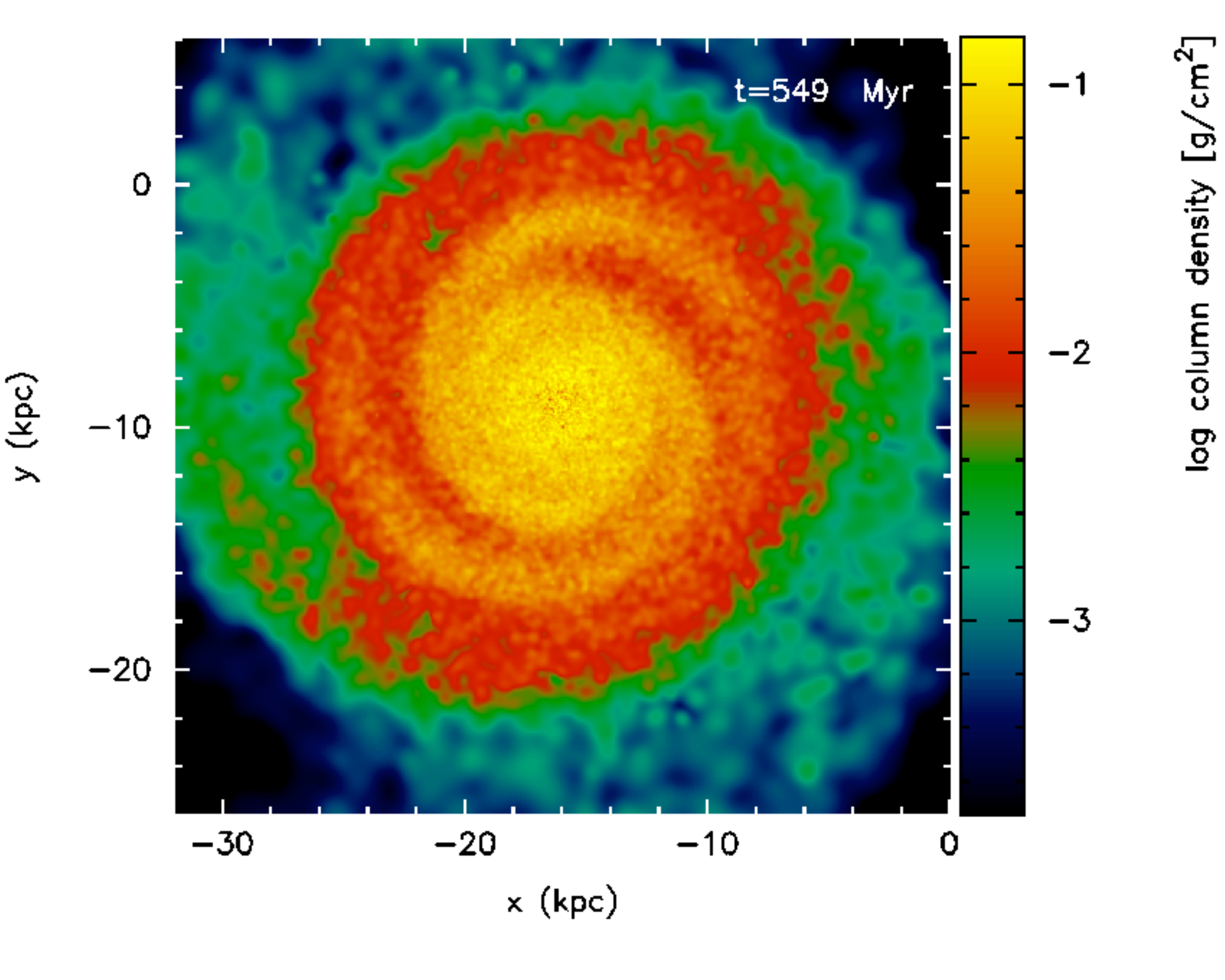}}
\centerline{
\includegraphics[scale=0.25]{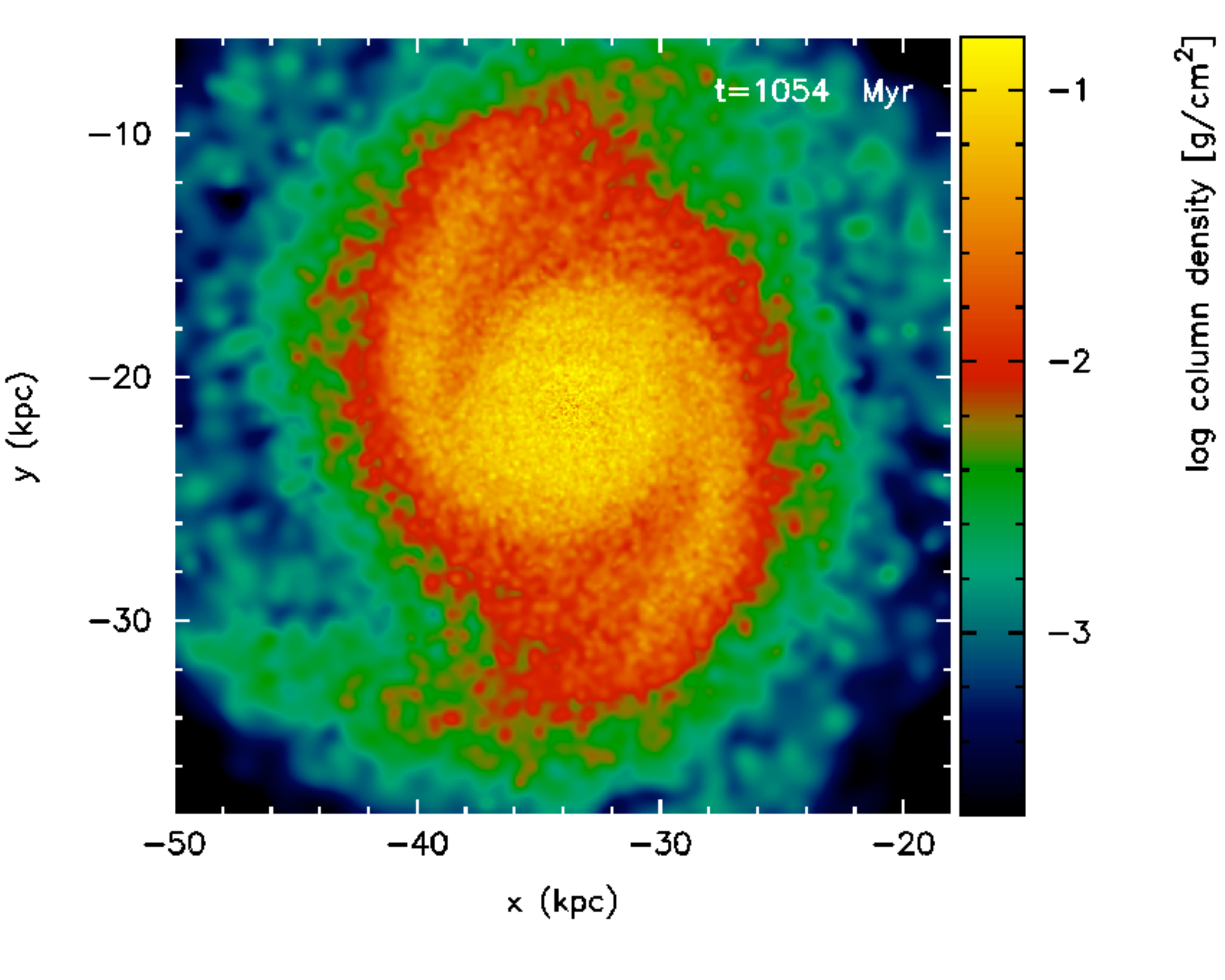}}
\caption{The stellar spiral arms are shown at times of 550 Myr and 1050 Myr. The stellar arms are relatively broad and weak. At 1050 Myr, the spiral arms are more difficult to discern at lower radii.}
\end{figure}
In Figures~4 and 5 we show the evolution of the gas column density for the 0.1 and 0.3 ratio companions. We also modeled a 0.01 ratio companion encounter.  In all cases, the companion galaxy passes below the primary galaxy (in the plane of the sky, or the $xy$ plane as shown in Figs.~3, 4 and 5) and then continues in the positive $x$ direction. The companion also moves from in front of to behind the main galaxy during its orbit.

For the 0.01 ratio companion, there is no immediate evidence of an
interaction as the companion passes the main galaxy. However the
companion galaxy slows substantially at this time, and gradually
exerts more influence on the main galaxy. By 380 Myr, there is a clear
tidal arm on the side of the companion, but not an obvious $m=2$
spiral structure. By this time, the companion is 50 kpc away as
seen on the plane of the sky, and about 100 kpc behind the main
galaxy, so a total overall distance of 112 kpc. This tidal structure
is relatively weak in nature and only lasts about 50 Myr (from 350 Myr to
400 Myr). 

For the 0.1 ratio companion, there is a much more obvious two armed
spiral structure. The $m=2$ pattern only occurs some time after
the companion has passed the galaxy, but is
still evident at a time of 900 Myr. At this point the companion lies
approximately 170 kpc away on the plane of the sky, and 200 kpc behind the
primary galaxy.

Finally when we take the 0.3 ratio companion, which is comparable to
the ratio between NGC 5195 and M51, we see a strong $m=2$
perturbation. This time, the perturbation is induced immediately as
the companion passes the galaxy, although it again becomes stronger
with time. The gaseous
spiral arms extend to the inner regions of the disc, although as
we discuss below, we cannot resolve the central parts of the gaseous
disc. With the 0.3 ratio companion, the spiral structure again persists
for the entirety of the calculation (1.15 Gyr), so lasts at least 900 Myr. At the end
of the calculation, the galaxies are 215 kpc apart on the plane
of the sky, whilst the companion lies 200 kpc behind the main galaxy,
thus a total distance of nearly 300 kpc. 

For both the 0.1 and 0.3 ratio companions, the $m=2$ structure
persists for the duration of our simulations. We would expect
eventually that larger order perturbations become dominant again at
some point (unless the companion induced the formation of a bar, but it
is not clear from our simulations that this is the case). We note that for a lower resolution calculation (with 50,000 stellar disc particles, see Appendix), the $m = 2$ structure did not persist for as long, only 750 Myr.

We also notice from Figures~4 and 5 that the extent of the spiral arms,
and even the gaseous disc, become smaller over time. This is
presumably a consequence of the spiral arms slowly winding up over
time (see below). 

Figures 6 and 7 show the larger scale distribution of gas after 550 or 1050 Myr. 
The distribution of gas also shows that the
perturbation is sufficient to produce very long, broad tidal arms of low
density gas.   
\begin{figure}
\centerline{
\includegraphics[scale=0.25]{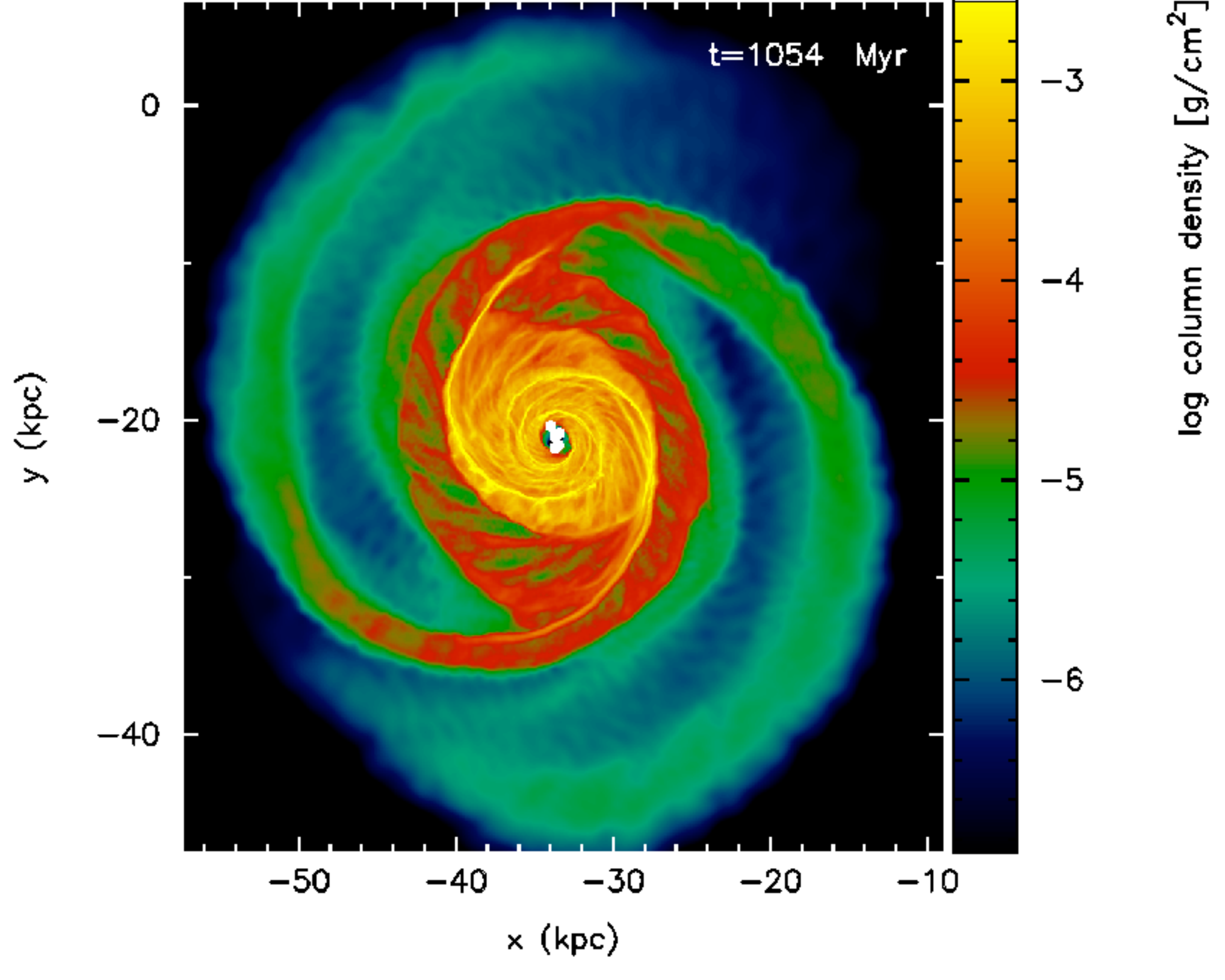}}
\caption{This figure shows the larger scale gas distribution at a time of 1050 Myr. The perturbation has led to long gaseous tidal arms extending to radii of 20 kpc.}
\end{figure}

In all our calculations, a build up of gas in the central regions
leads to high densities and subsequently the formation of sink
particles. With a strong spiral perturbation, there is also inflow of
gas to the centre. As the sink particles orbit the centre, all the gas
within their orbit becomes gradually accreted and a void
develops at later times (this is much less noticeable in the isolated
case). Thus it is impossible to resolve the gas in the inner parts of
the galaxy. However the gas mass is very low, so the dynamics of
the centre should not influence the evolution of the galaxy.       

\subsubsection{Distribution of stars}
In Figure~6 we show the distribution of stars for the 0.3 ratio
companion at times of 550 and 1050 Myr, approximately 400 and 900 Myr
after the interaction. The top panel shows that the stellar spiral
arms extend to the inner parts of the disc, as much as our simulation
permits. However we are limited by resolution in the inner part of the
disc, particularly since the stellar arms are so tightly wound.   We note as well that although the perturbation in the stars is
relatively weak and broad. Moreover, there is a good deal of smoothing in the stellar waves, presumably due to the epicyclic motions of the stars, including a pre-collision radial component to their motion. Figures~4 and 5 indicate that there is a very sharp spiral arm in the gas as it shocks, similar to
\citet{Dobbs2010}. The gaseous spirals are also similar to the less smoothed stellar spirals of Set 1, shown in Figure~1.

At the later time of 1050 Myr, we still see spiral structure in the
stars but it appears to be limited to large radii. This probably
indicates that the inner stellar arms have largely wound up in the
inner 4 or 5 kpc (as in the models of Set 1) and either there is little spiral structure, or it is
too tightly wound to resolve. The gas still
retains an $m=2$ spiral pattern, even in the inner parts.

\subsubsection{Evolution of Fourier modes}

\begin{figure*}
\centerline{
\includegraphics[scale=0.3]{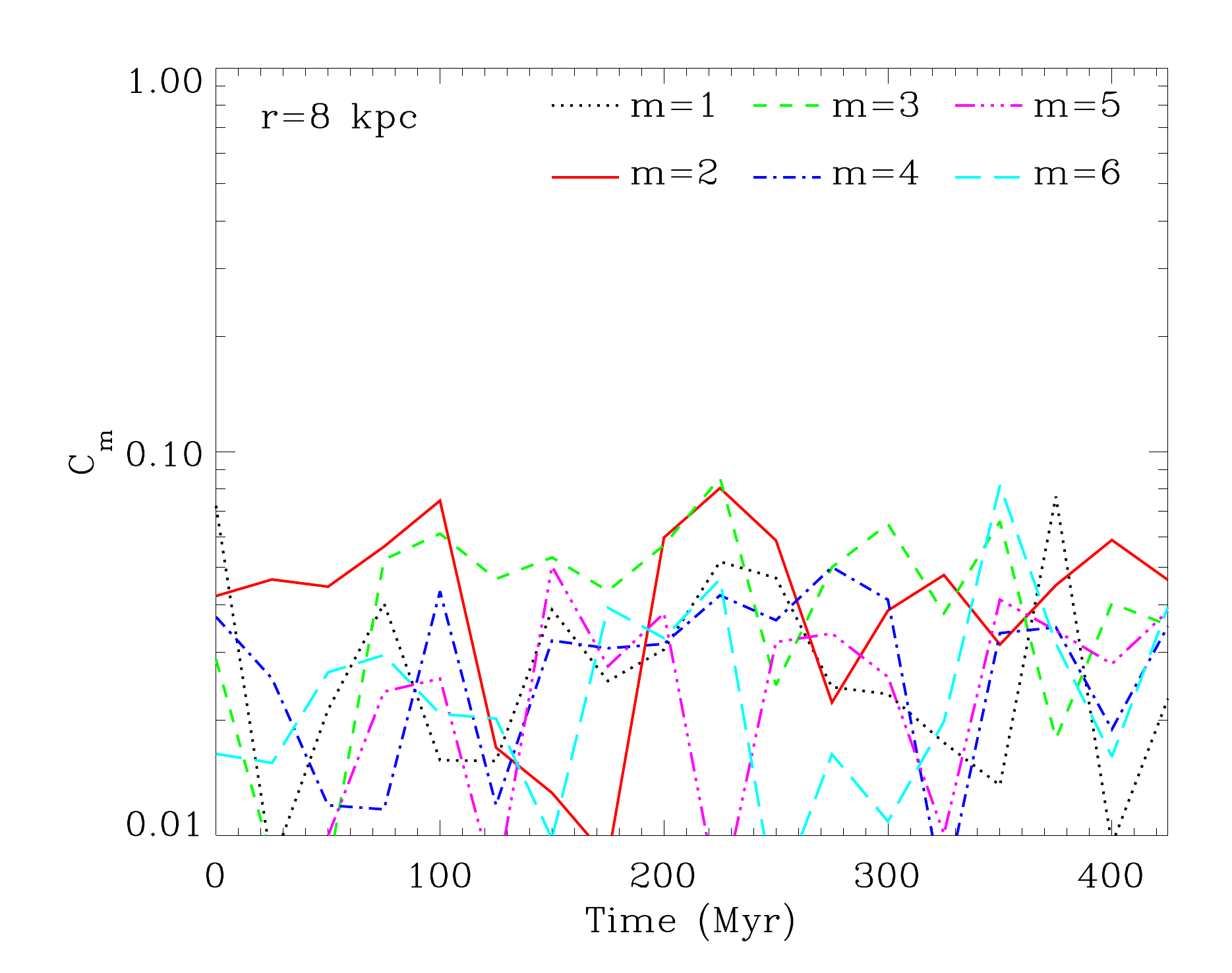}
\includegraphics[scale=0.3]{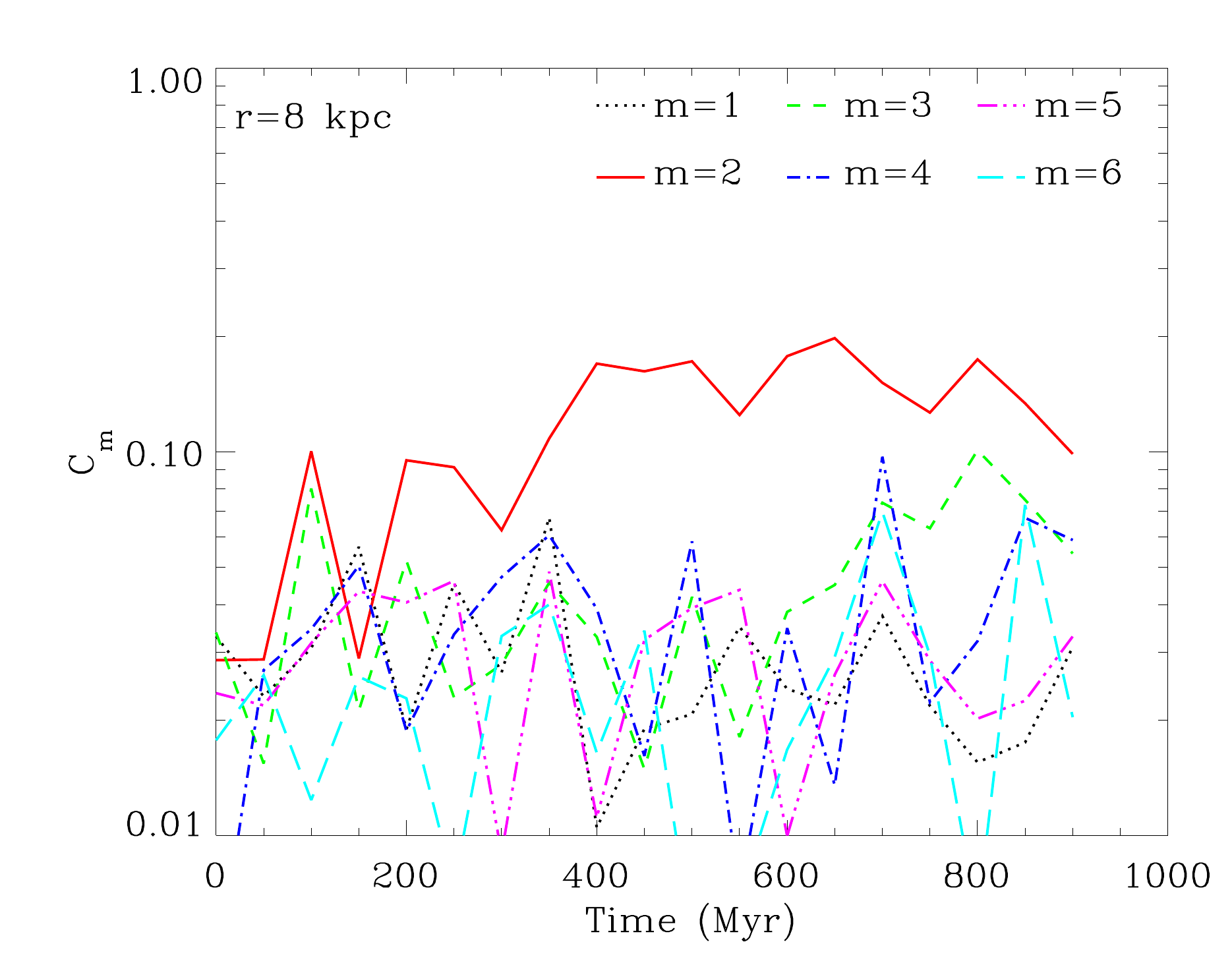}
\includegraphics[scale=0.3]{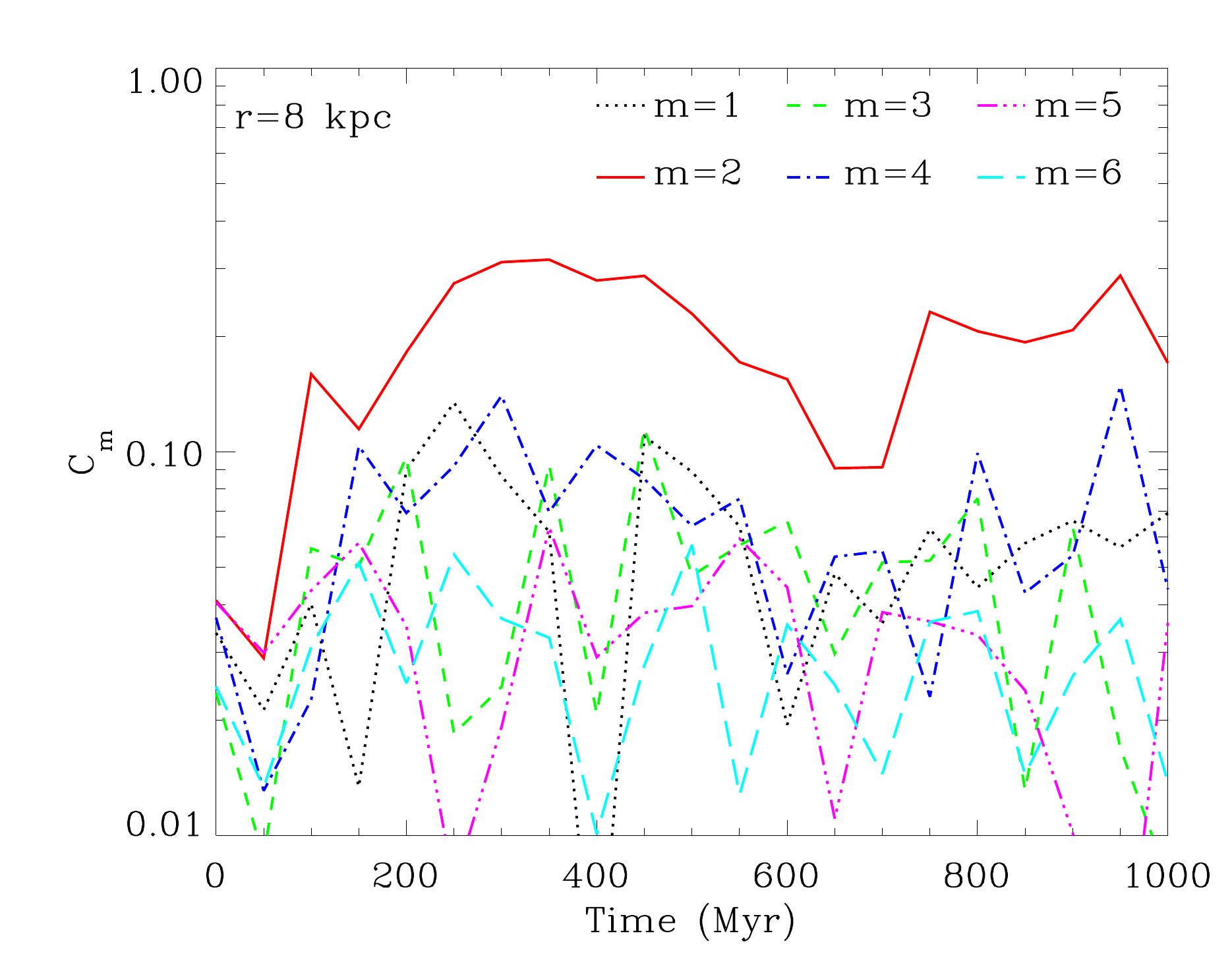}}
\caption{The Fourier amplitudes of the stellar disc for modes $m=1-6$ are plotted versus time for the 0.01 (left), 0.1 (middle) and 0.3 (right) ratio companion. The $m=2$ mode becomes dominant for the 0.1 and 0.3 ratio companions, but there is no evidence for an $m=2$ perturbation from the Fourier modes for the 0.01 ratio companion.}
\end{figure*}  

We calculate the Fourier modes of the galaxy in the same way as presented in \citet{Dobbs2010}, according to 
\begin{equation}
  C_m=\frac{1}{M_{\rm disc}} \bigg|\int^{2 \pi}_{0} \int^{R_{\rm
      out}}_{R_{\rm in}} \Sigma(R,\theta) R \, dR \thinspace 
  e^{-im \theta} d\theta \bigg|,
\end{equation}
where $M_{\rm disc}$ is the mass of the stellar disc, $\theta$ is the
azimuthal angle and
$\Sigma$ is the corresponding surface density \citep{Theis2004}. 
We show the Fourier modes versus time in Figure~8 for the 0.01 (left),
0.1 (centre) and 0.3 (right) companions. The Fourier modes are
calculated for the stellar component of the disc. There is no dominant mode at any time for the 0.01 companion, so even though there is a visible influence of the companion on the structure, this is not perceptible in the Fourier components. For the 0.1 and 0.3 ratio companions, there is a clearer dominance of the $m=2$ mode, although it is often only a factor of 2 or 3 higher than the other modes. Again the $m=2$ mode is stronger at relatively earlier times for the 0.3 ratio companion.  

The $m=2$ mode is highest down to radii of about 5 and 7 kpc for the
0.3 and 0.1 ratio companions (Figure~9). However, Figure~6 shows that the
stellar spiral arms are very tightly wound at smaller radii, and given
that they are relatively broad, it may be difficult to extract the
spiral arms with our resolution. 

\begin{figure}
\centerline{
\includegraphics[scale=0.3]{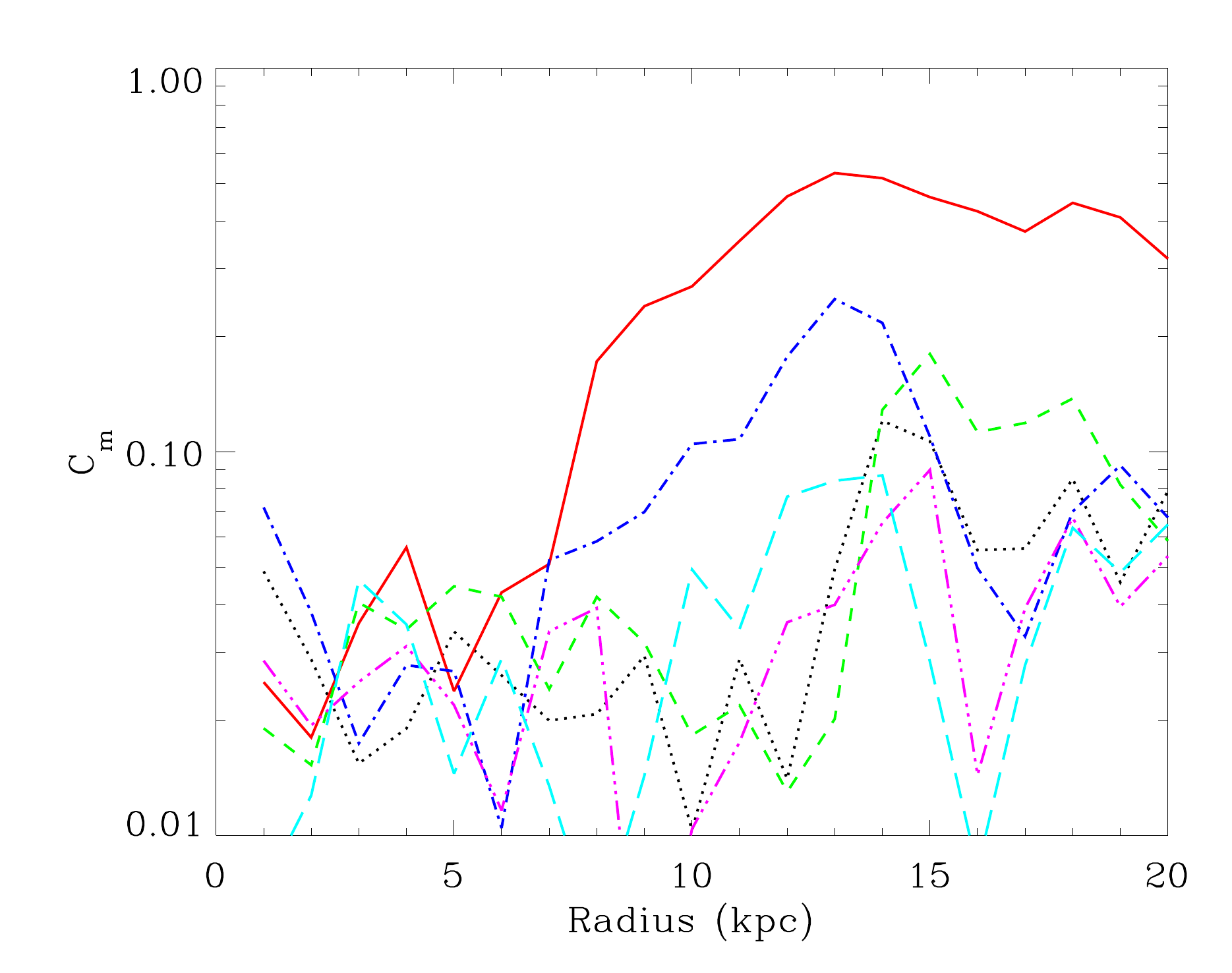}}
\centerline{
\includegraphics[scale=0.3]{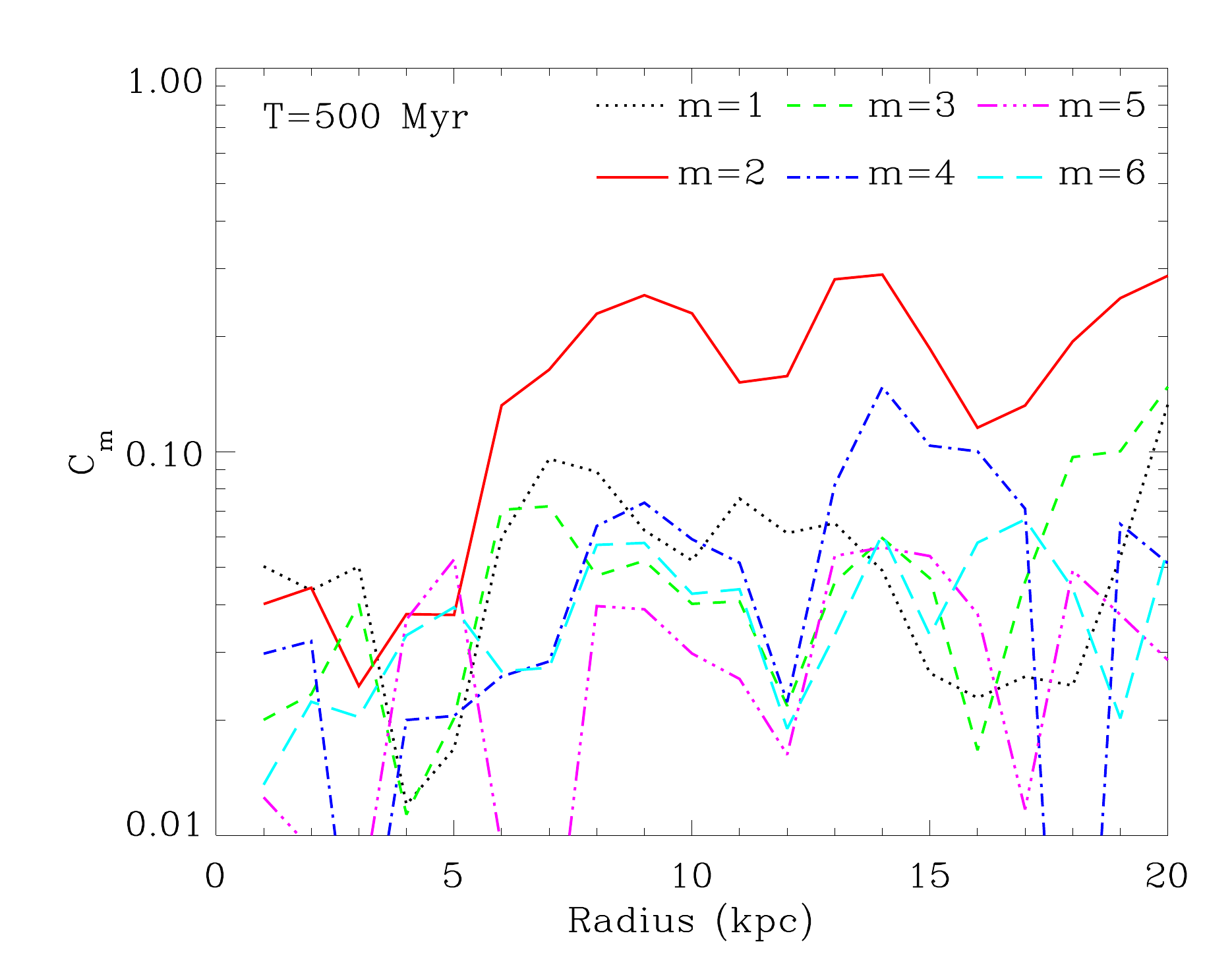}}
\caption{The Fourier amplitudes of modes $m=1-6$ are plotted 
  versus radius for the 0.1 (top) and 0.3 (lower) companions. The
  $m=2$ mode is dominant down to radii of 7 kpc for the lower mass
  companion, and 5 kpc for the 0.3 ratio companion, though this
  difference may not be significant. Furthermore the $m=2$
  perturbation may extend further inwards with higher resolution
  calculations, since the spiral arms become very tightly wound at small
  radii.}
\end{figure}  

\subsubsection{Evolution of arm strength}
An alternative measure of the response of the disc to the tidal perturbation of the companion galaxy is the arm strength, $F$. This can be defined as 
\begin{equation}
F=\frac{2 \pi G \delta \tilde{\Sigma}_{m=2}}{R \Omega^2}
\end{equation}
where $\delta \tilde{\Sigma}_{m=2}$ is the amplitude of the $m=2$ perturbation \citep{Oh2008}. $F$ gives a measure of the ratio of the gravitational force perpendicular to the arms to the unperturbed case. We show the variation of $F$ with time (upper panel) and radius (lower panel) for the 0.3 and 0.1 ratio companions in Figure 10. The strength $F$ is calculated for the stars; $F$ will be minimal for the gas owing to the low gas surface density assumed in the models. We see that $F$ is typically 10-20 per cent for the 0.3 ratio companion, and $\lesssim 10$ per cent for the 0.1 ratio companion. In both cases, $F$ stays relatively high for the duration of the calculation. $F$ peaks later, and at larger radii for the 0.1 ratio companion.  
\begin{figure}
\centerline{
\includegraphics[scale=0.3]{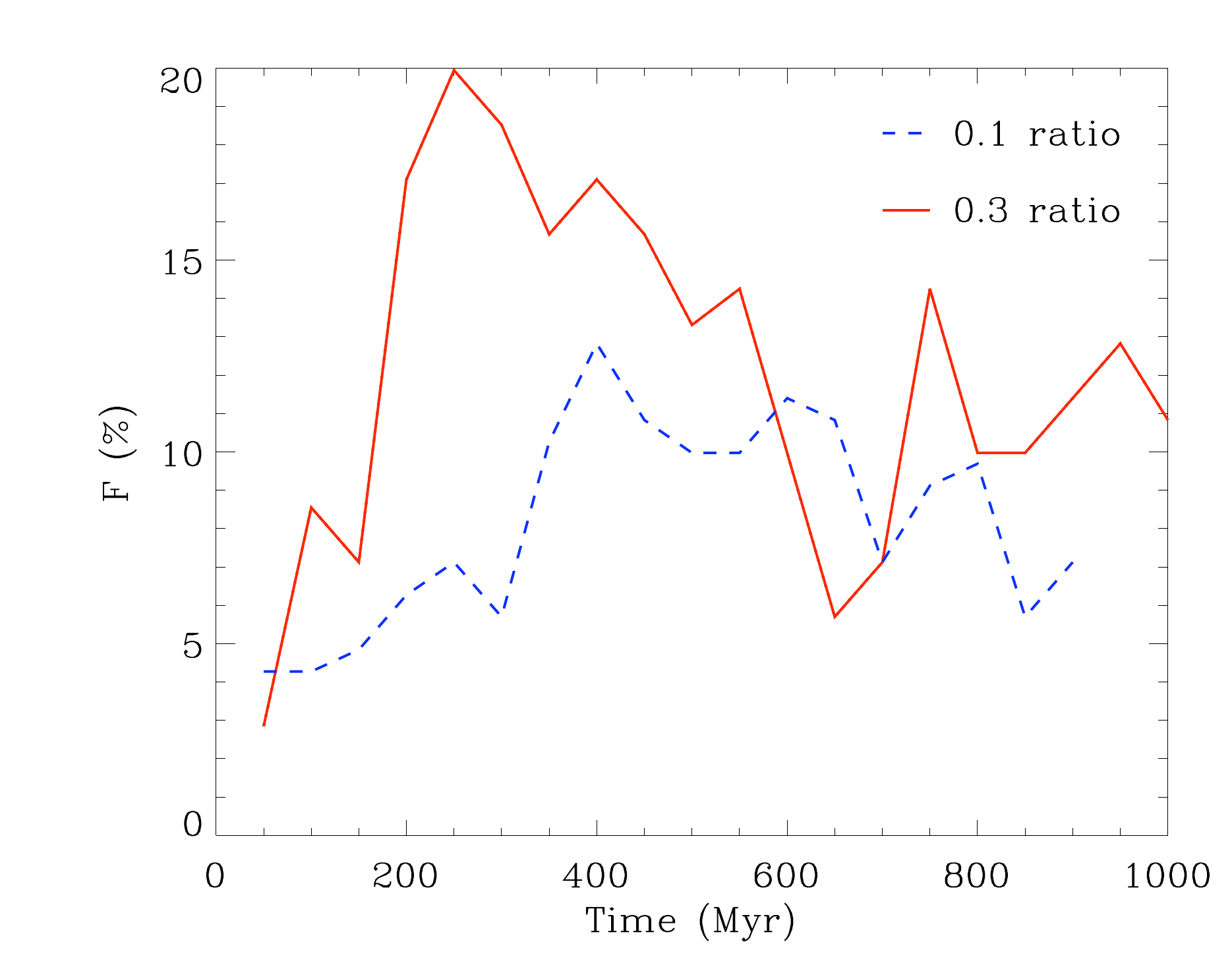}}
\centerline{
\includegraphics[scale=0.3]{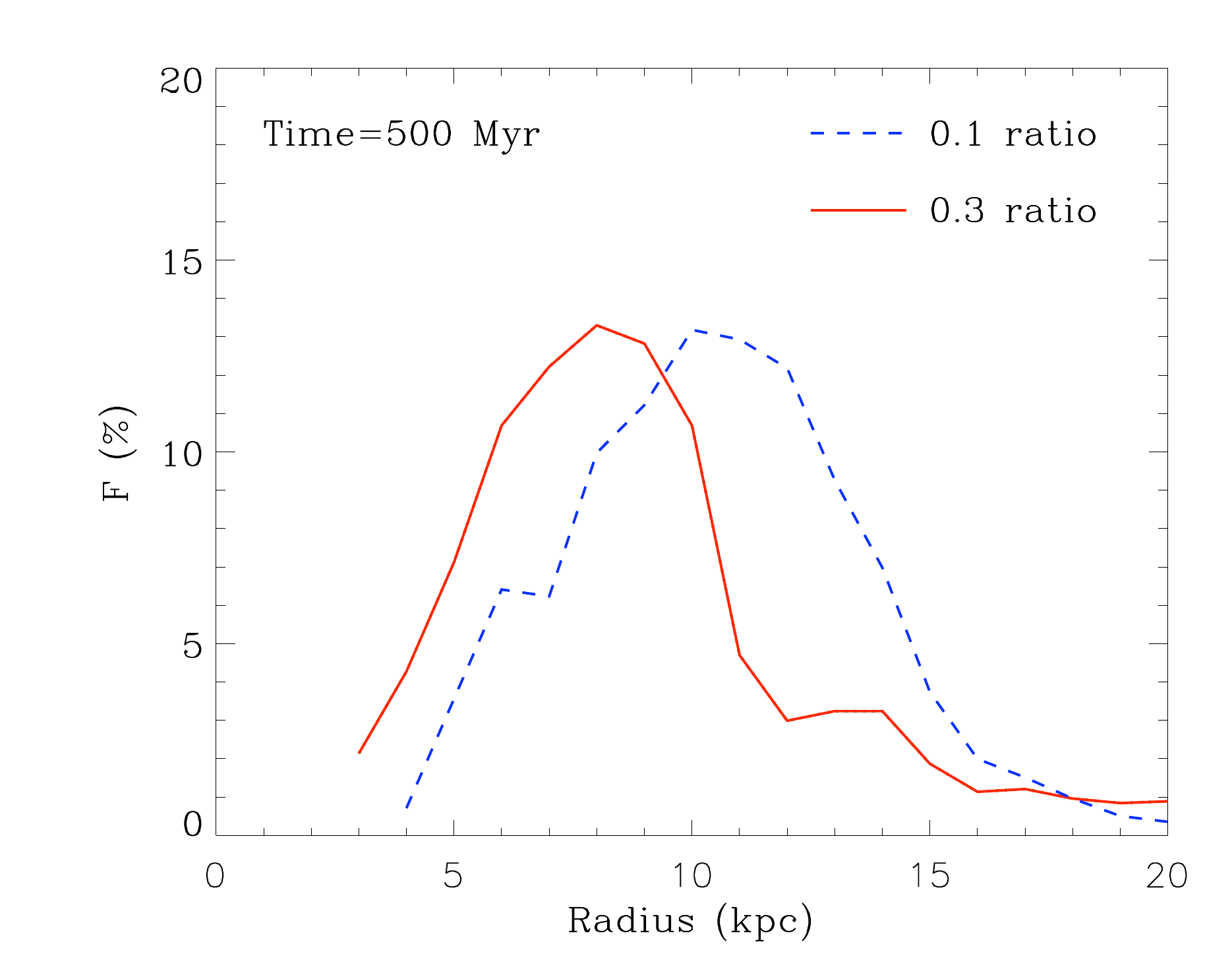}}
\caption{The arm strength, $F$, is shown versus time (top) and radius (lower panel). $F$ is calculated for the stars, for the models with the 0.1 and 0.3 companions. The arm strength is of order 10 per cent for both cases. For the top panel, $F$ is calculated at a radius of 8 kpc, whilst in the lower panel, $F$ is calculated at a time of 500 Myr.}
\end{figure}  

\subsubsection{Offset between stars and gas}
\citet{Dobbs2010} found that for simulations designed to model M51, there was no measurable offset between the stars and gas. Here we take a similar approach, fitting the gaseous and stellar arms to a Gaussian function of the form
\begin{equation}
  \rho(\theta)=A_1 \exp \bigg[-\Big(\frac{\theta-B_1}{C_1}\Big)^2 \bigg]+ A_2 \exp \bigg[-\Big(\frac{\theta-B_2}{C
_2}\Big)^2 \bigg]+A_3,
\end{equation}

\noindent
with the amplitudes of the spiral arms given by $A_1$ and $A_2$, the
offsets by $B_1$ and $B_2$ and the dispersions by $C_1/\sqrt{2}$ and
$C_2/\sqrt{2}$. Here we analyse the simulation where the companion ratio is 0.3, since this produces the strongest perturbation. 

In Figure~11 we show the location of the stellar and gaseous arms at times of 350 (top) and 550 (lower) Myr.
At neither time is a consistent offset evident between the stars and gas. Instead, the gas and stars are largely coincident, which is not surprising given that this was the case in our previous calculations of M51 \citep{Dobbs2010}. There is a larger discrepancy between the stars and gas at 550 Myr, but the gas is located both down and upstream of the stars at different radii. The large difference at 10 kpc reflects that the stellar arms end roughly at this radius. 

\begin{figure}
\centerline{
\includegraphics[scale=0.4]{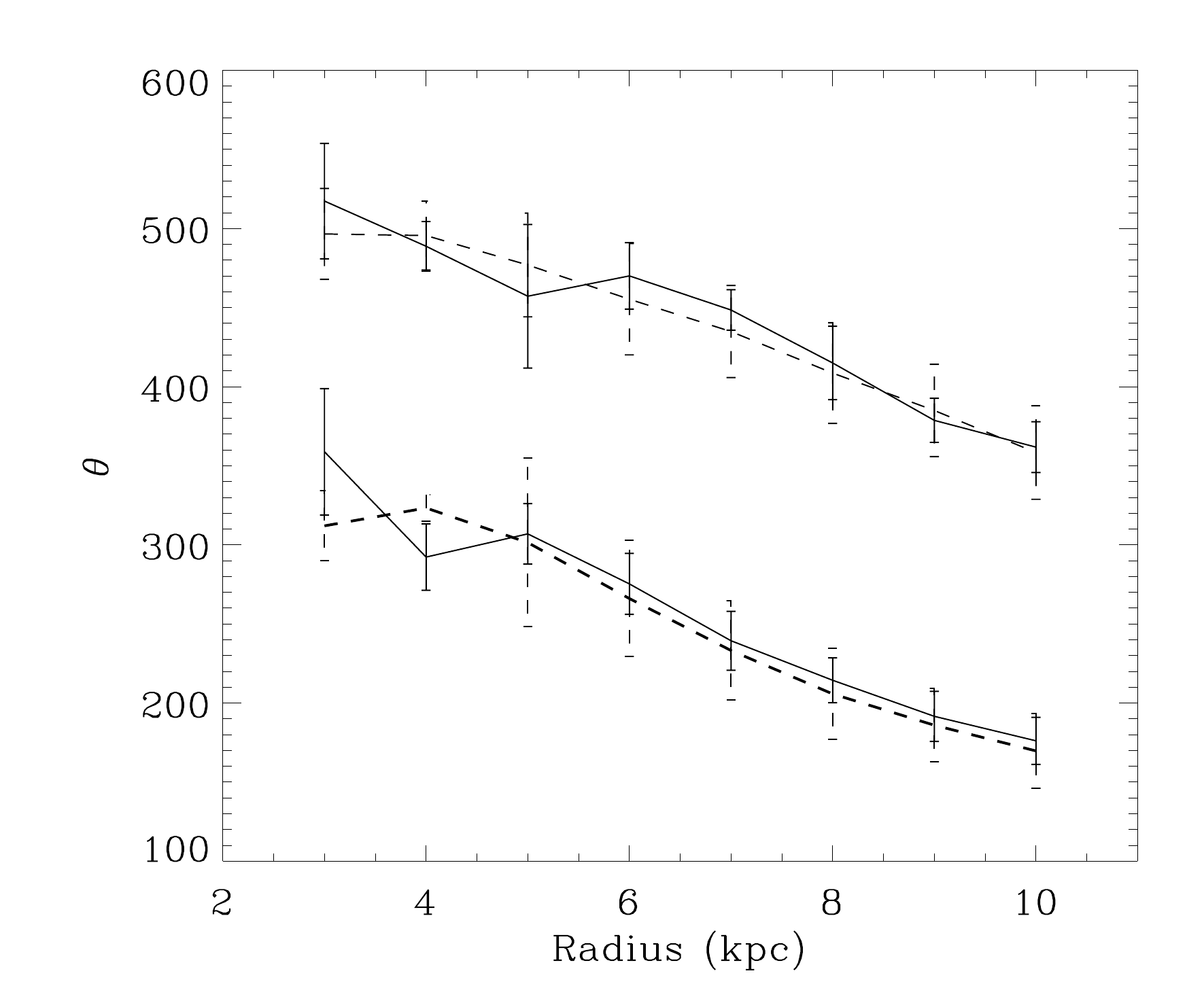}}
\centerline{
\includegraphics[scale=0.4]{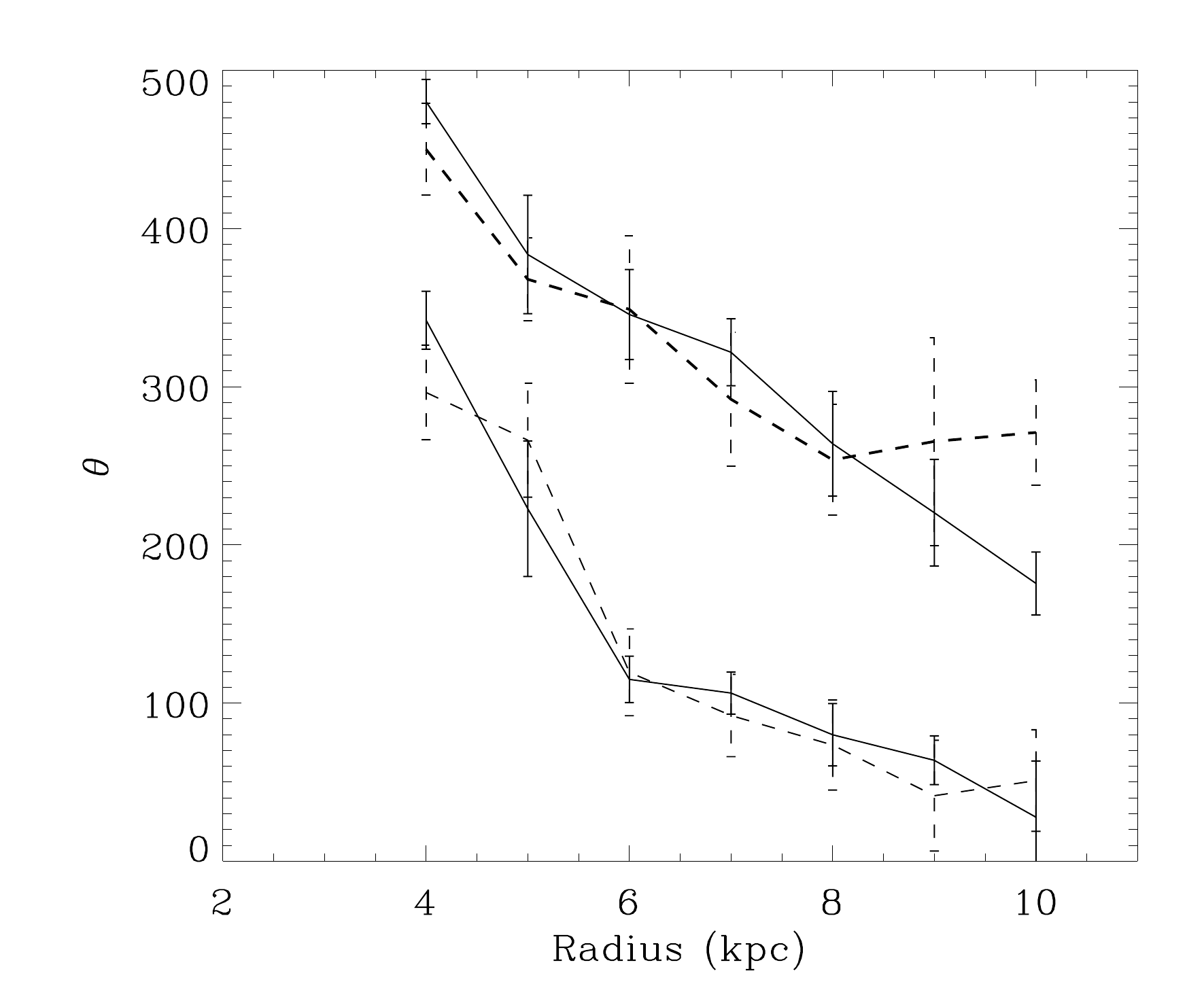}}
\caption{The location of the gaseous (solid lines) and stellar (dashed) spiral arms are shown after 350 Myr (top) and 550 Myr (lower) for the model with the 0.3 ratio companion. At neither time is there a noticeable offset between the two components.}
\end{figure}  

\subsubsection{Pattern speeds}
We can tell from Figures~5 and 11 that the spiral arms wind up after the interaction. We show the pattern speed versus radius in Figure~12 using the locations of the gaseous and stellar arms at 350 and 550 Myr. Our calculation of the pattern speed is hindered by the difficulty in determining the location of the spiral arms in the inner parts of the disc. Nevertheless, the pattern speed almost exactly follows the inner Lindblad resonance, only offset by about 10 km s$^{-1}$. As found in \citet{Dobbs2010}, the spiral arms most resemble kinematic density waves. The pattern speed is higher than that predicted for purely kinematic density waves (i.e. $\Omega-\kappa/2$) due to the self gravity of the stellar disc, which reduces the rate at which the pattern unwinds.

Our analysis does not assume that the spiral arms in our models are logarithmic.  Figure 11 indicates that at 350 Myr (top panel), the pitch angle is reasonably constant with radius, and $\tan i \sim 0.24$. At the later time of 550 Myr, though the error bars are large, the arms appear more irregular, as can also be seen in the lower panels of Figure 5. Overall the average pitch angle decreases with time according to $\tan i \propto t^{-0.8}$, although the rate of change decreases with time. \citet{Oh2008} find that $\tan i \propto t^{-0.6}$  and $\tan i \propto t^{-1}$ for self-gravitating and non-self gravitating calculations respectively.    
\begin{figure}
\centerline{
\includegraphics[scale=0.4]{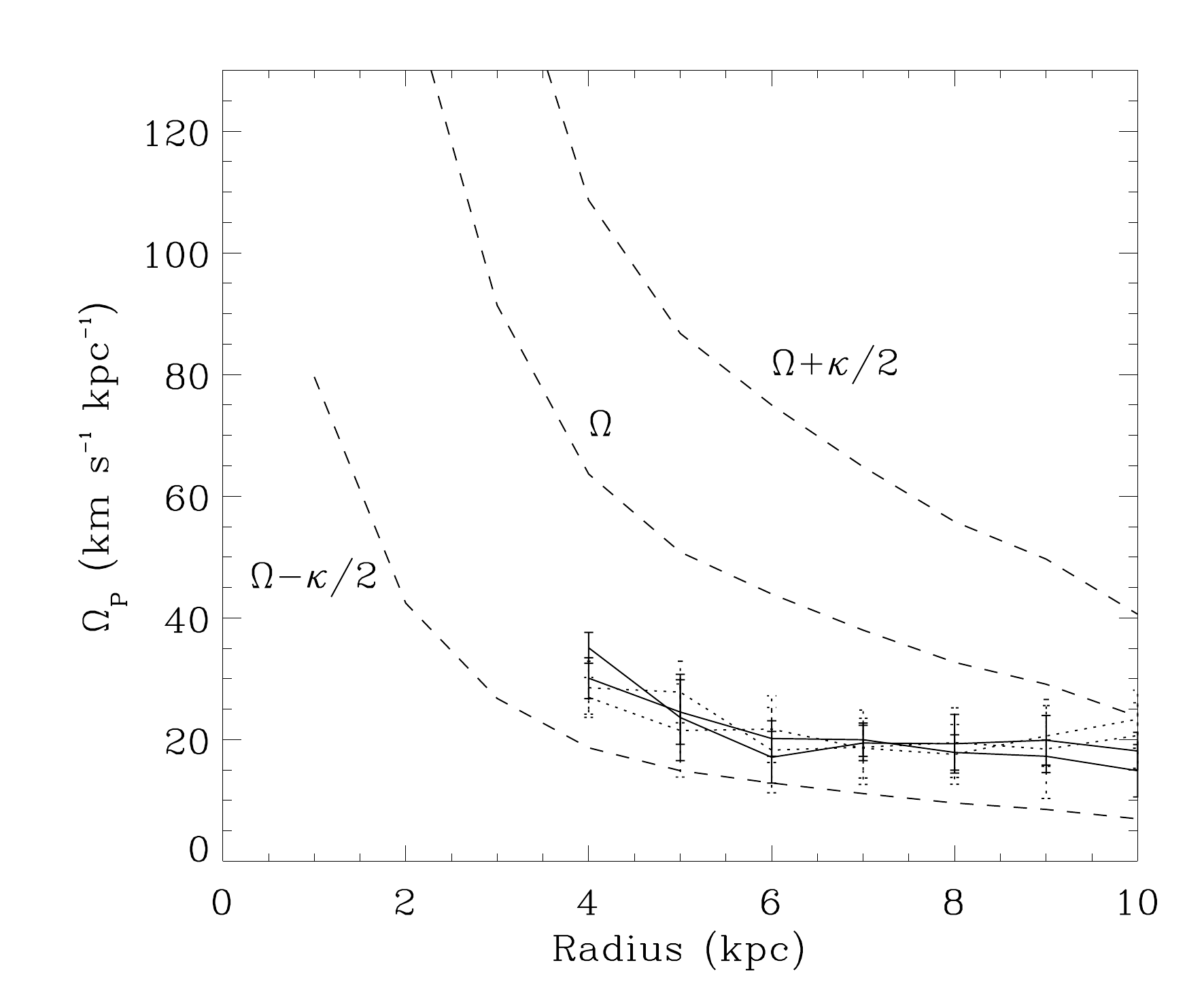}}
\caption{The pattern speed is plotted for the model with the 0.3 ratio companion, indicating that the spiral pattern is slowly winding up, at a rate slightly faster than the inner Lindblad resonance. The solid lines represent the gaseous arms, and the dotted the stellar arms.}
\end{figure} 

\section{Comparisons to Analytic Models}
\subsection{The Analytic Formalism}

To understand the results of the numerical simulations we analytically calculated the orbits of stars initialized on a disc-covering grid. Assuming a perfectly flat rotation curve, for simplicity, each star was initially assigned an azimuthal velocity equal to the circular velocity, and zero radial velocity. In the Impulse Approximation it is assumed that in the encounter each star experiences an acceleration $a$ in the direction of the companion for a short time interval ${\Delta}t$. In our analytic model we assumed that the $a$ equals the classical, textbook tidal acceleration, which is also the small amplitude limit of the complete expressions (see e.g.,  \citealt{ge94}). Specifically, we add the following velocity component impulses to the velocities of each star:


\begin{equation}
\Delta{v_r} = {\Delta}V \frac{r}{D}cos(2\phi), 
\ \ 
\Delta{v_{\phi}} = {\Delta}V \frac{r}{D}sin(2\phi), 
\end{equation}

\noindent
where ($r, \phi$) are the radial and azimuthal positions of the star, $D$ is the total radius of the disc, and ${\Delta}V$ is a constant velocity amplitude factor that includes the usual tidal constants and the ratio of the disc size to the distance of closest approach. 

Then the epicyclic approximation is used to compute the stellar orbits. Specifically, the orbit equations are



\begin{eqnarray}
r(t) & = & q - Aqsin \left( {\kappa}t + {\psi} \right), \\
\phi(t)&  = & \phi_o + {\omega}_{cir} (q)t + \nonumber \\
&&\sqrt{2} A \left(  cos \left( {\kappa}t + {\psi} \right) - cos(\psi) \right),
\end{eqnarray}

\noindent
where ($q, \phi_o$) are the initial values of ($r, \phi$), $\kappa$ is the epicyclic frequency, $\omega_{cir}$ is the circular frequency, $\psi$ is the epicyclic phase after the impulse, and $A$ is the epicyclic amplitude. Note that because of the angular momentum change resulting from the azimuthal velocity impulse, the radius of the guiding centre circle (i.e., of the epicycle's centre) does not equal the star's initial radius. In these equations, $q$ equals the former not the initial stellar radius, and all of the quantities: $\kappa$, $\psi$, $\omega_{cir}$, and $A$ are dependent on this $q$, which can be derived from the initial radius and azimuthal impulse. The values of $\psi$ and $A$ in terms of  ${\Delta}V$ and the initial positions are derived by comparing the epicyclic velocities derived from equations (5) and (6) to those of equation (4). These calculations are described in detail in Appendix 2 of \citet{ap96}.

\subsection{Comparison of the Semi-Analytic Models to Set 1 Simulations}

Figure 13 shows an example of stellar disc evolution according to this prescription. Specifically, a perturbation amplitude of ${\Delta}V = 0.1$ was chosen. The analytic trajectories, and thus the disc evolution, are then calculated as described in the previous subsection. The figure shows snapshots at three times, as described in the caption, with the final panel in the lower right showing a snapshot of the numerical model of Figure~1 (Set 1) for direct comparison with the analytic model. In order to make the comparison two adjustments were made to the analytic model disc. First, it was rotated by 90$^\circ$ to better match the point of closest approach in the numerical model. Secondly, the time was adjusted by subtracting an interval of 0.45 units. The reason for this offset is that the velocity impulse is applied instantaneously in the analytic case, but is not realized in the numerical model until the companion is well past closest approach. 

\begin{figure*}
\centerline{
\includegraphics[scale=0.45]{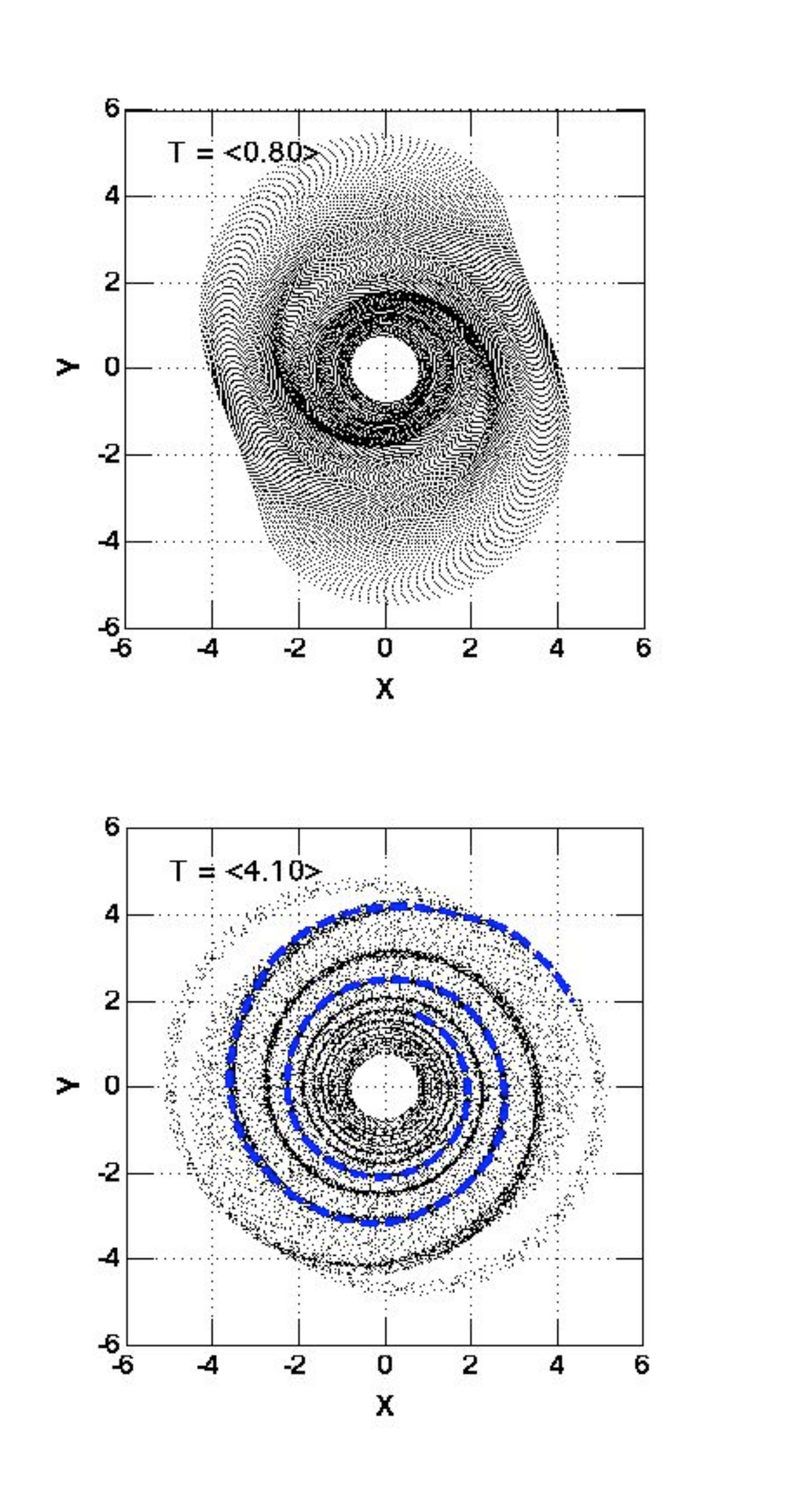}
\includegraphics[scale=0.45]{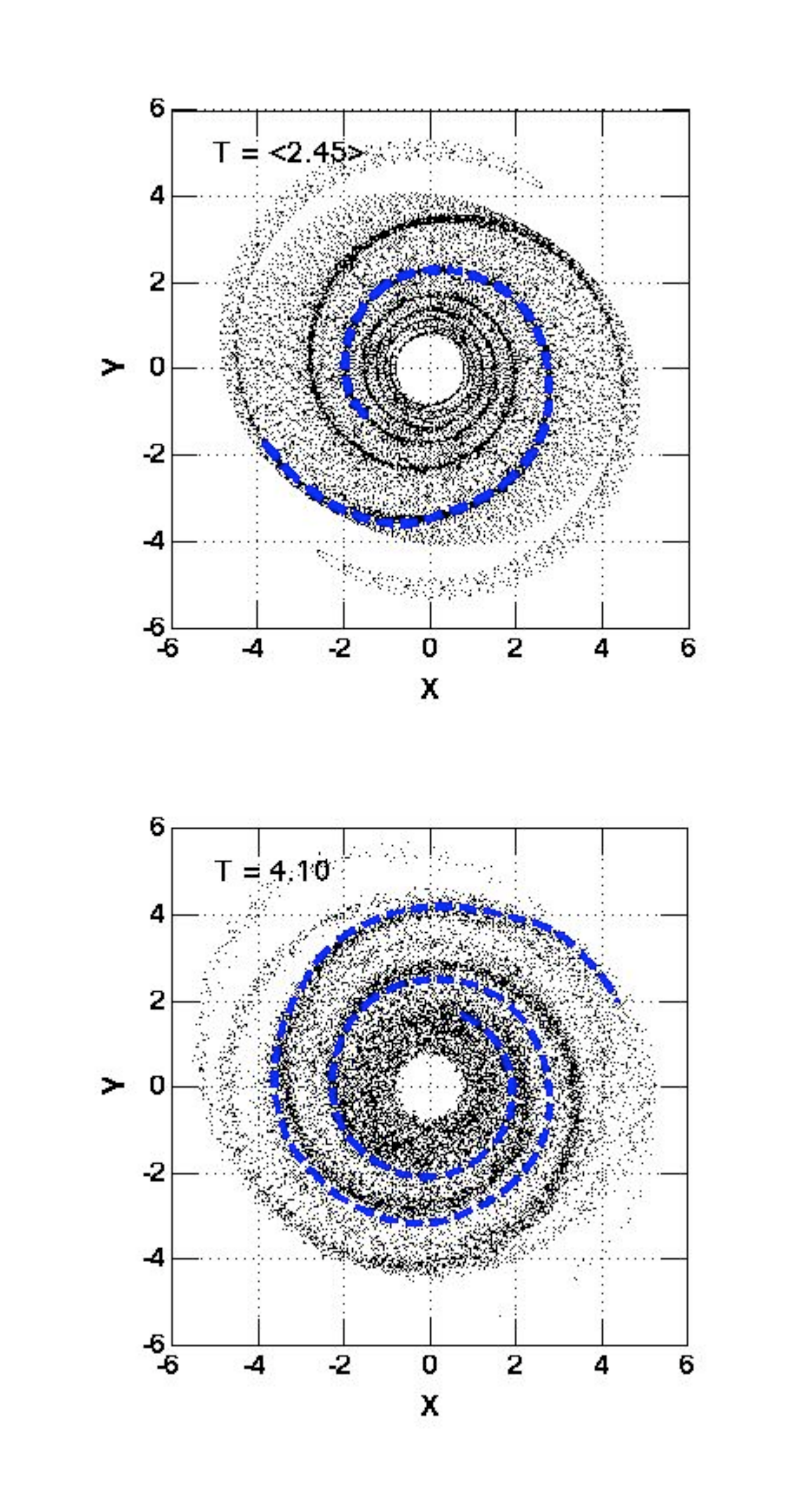}}
\caption{The first three panels display the analytic model at times corresponding to those of the numerical model shown in Figure~1. The dashed curves (blue in the electronic edition) in the last three panels show the analytic caustic centre condition superimposed on one of the two symmetric spirals. The fourth panel shows the numerical model of Figure 1 at the same time as the analytic model in the third panel. The caustic centre curve, which is the same in both third and fourth panels, confirms that analytic and numerical models are nearly identical in the outer disc, but differ in the inner disc, where the gravitational potentials are different.}
\end{figure*} 

With these adjustments, the comparison in the bottom row of Figure 13 is very good. Comparisons between the panels in the top row and the corresponding timesteps in Figure 1 are also good. The most obvious difference is that the spirals are discernable to very small radii in the analytic, but not the numerical model. Experimentation shows that this is a function of the amplitude $A$ in the analytic model, but we have made no attempt to optimize that fit. It is also the result of different halo potentials in the central regions. The numerical model has a rising potential in the inner half of the disc, not a flat one like the analytic model. The curve drawn over one of the two waves in the last two analytic panels and transposed from the last onto the numerical model, fits in the outermost parts, but diverges in the inner. 

Those curves, which are meant to represent the wave centre, are derived in the same way as in the ring wave theory. In that application, and in the case of more asymmetric waves, the locations of caustics are given by the Jacobian matrix whose elements are derivatives of the coordinates of equations (5) and (6) with respect to their initial values $q$ and $\phi_o$ (see \citealt{ap96}, \citealt{sm90}, and \citealt{ge94}). Moreover, the birthplace of caustics and the centre of paired caustics, which are the inner and outer wave edges, are given by a matrix of second derivatives. In the present case, the cross terms in those derivative matrices are negligible (e.g., ${\partial}r/{\partial}{\phi_o}$ is small), so as in the case of ring galaxies we require ${\partial}r/{\partial}q = 0$, and similarly for the second derivative. The latter condition yields an equation identical to equation (4.14) of \citet{ap96}, except for the addition of the phase ${\psi}$. (It is the variation of this phase that changes a ring wave to a spiral.) The dashed curves drawn in Figure 13 derive from this equation. In the lowest order approximation this equation has the form, 

\begin{equation}
\phi_o = n\pi - {\kappa}(q)t/2, 
\end{equation}

\noindent
where $n$ is an integer and ${\kappa}(q) \sim 1/q$ for a flat rotation curve. The ($r, \phi$) expression for the spiral can be obtained by using this expression in equations (5) and (6). Specifically, the procedure is as follows. Select values of time $t$ and initial azimuth $\phi_o$, and use the above equation to derive a value of the initial radius $q$. In the present context $q$ is essentially a label telling us which particle had the given initial azimuth, and is located at the spiral centre at the given time. The amplitude $A$ and the initial particle phase $\psi$ in equations (5) and (6) are determined with the aid of equation (4) given the velocity impulse $\Delta V$ and the derivative of equations (5) and (6). 

\subsection{Comparison to Set 2 Models}

Figure 14 shows a comparison between an N-body, SPH model from Set 2, and a scaled version of the analytic spiral centre curves described in the previous subsection. The fitting procedure was as follows. An analytic curve chosen at a roughly corresponding time, was translated, rotated and expanded to fit the numerical output at t = 575 Myr. This output is in the middle of the numerical run and has well-developed spirals, but is otherwise arbitrary. Given the many adjustable variables, the good fit at one time is not too significant. Next an analytic curve at an earlier time was chosen to fit the output at t = 254 Myr, with the same translation, rotation and expansion factors. These two fits allow us to derive the time transformation coefficients between the analytic and numerical models. Specifically, we assumed that time in the analytic model was linearly related to that in the simulation with a scale factor to account for different units, and a constant delay in the analytic model, as described in the previous subsection. E.g., $t_{anal} = at_{sim} +b$. We then used the time values at the two selected instants to solve for $a$ and $b$. With this, we derived the corresponding times in the analytic model for the other outputs shown in Figure 14, and superposed the corresponding analytic curves. In the case of the last output, at t = 1141 Myr, the fit is beginning to break down in the outer disc. 

Despite that, it is clear from Figure 14 that the analytic curves fit the numerical spirals well over the selected azimuthal range, which usually extends for at least $180^{\circ}$ over a clearly defined spiral segment, and some additional less well defined parts of the wave. The analytic curves have not been extended very far into the central regions because the smoothness of the particle distribution in the simulation, and the different rotation curves make comparisons impossible there. As in the comparison to Set 1 models, we have not attempted to optimize the fit, e.g., by using the rotation curve of the initial numerical disc in the analytic model. In Figure 14, the analytic curve is also truncated at an arbitrary point in the outer disc where the model spiral is disappearing. Some divergence can be seen from the analytic curve in the outermost parts at all but the earliest times. Possible reasons for this include the simplicity of the fitting procedure, or the fact that in the numerical model the tidal disturbance is stronger there. Furthermore, although the spiral waves behave predominantly as kinematic density waves in these models, self-gravity is not negligible.

Some kinks can also be seen in the analytic curves. These are the results of mildly discontinuous jumps in the analytic solution from one spiral segment to the next, where these segments would correspond to the separate rings in a ring-making collision. Here the rings are broken and join to each other, but not completely smoothly at the current level of approximation. 

The fact that overall fits are good shows that even though the model spirals don't look especially tightly wound at the later times in the figure, they are indeed wrapping up. Evidently, the relatively open appearance is the result of only resolving the wave over a limited range of radii, and especially not in the centre where the tight-winding would be most apparent, if not for smoothing. 

\begin{figure}
\centerline{
\includegraphics[scale=0.40]{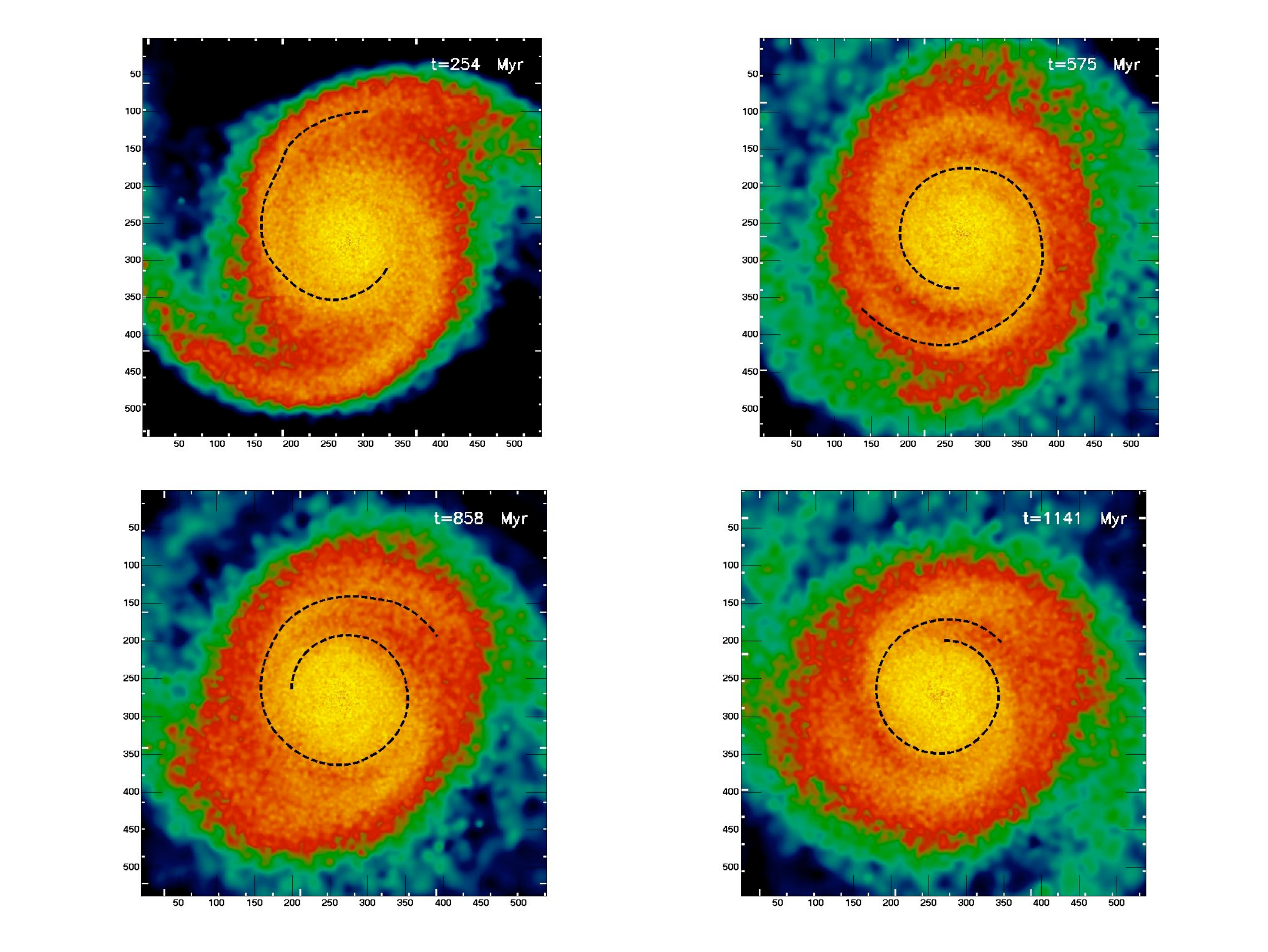}}
\caption{Selected snapshots of the distribution of stars in the Set 2 simulation described in Section 3 are shown. In each panel a dashed curve representing an analytic calculation of the wave centre is superimposed on one of the two spirals. The procedure for calculating these curves is described in the text.}
\end{figure} 

\subsection{Interpretations}

The agreement between numerical and analytic waves assures us that the persistance of the former is not a numerical artifact and confirms the nature of these waves. The question remains, however, of why do they not dissipate more rapidly as they wind up? In the context of density wave theory we might expect that as the waves shear their amplitude will decrease, unless swing amplification intervenes. However, these waves are not produced by a local disturbance that simply stretches away. They are the result of a globally correlated initial disturbance. That statement also applies to the original swing amplified wave simulations too (see \citealt{Toomre1981}). However, the effects of self-gravity were stronger in those model discs; evidently strong enough to dominate the effects of the perturbation correlations. 

Specifically, self-gravity can change the epicyclic phases of a set of orbits, allowing the waves containing them to damp. The Q parameter consists of the product of two ratios of a kinematic timescale (e.g., shear and sonic) and a local free-fall time. When local self-gravity is relatively weak (large Q), the waves are only damped on the phase mixing (or gaseous dissipation) timescale. (The relationship between swing amplification and caustic wave persistence is likely to be complex in discs with Toomre Q values near unity, and we will not attempt to understand it here.)

The sharp edges of the waves visible in both our analytic and numerical models provide the clue to their nature. These are caustic edges formed by strong orbit crowding at the inner and outer edges of an orbit-crossing zone in a wave. This is shown clearly in Figure 15, which plots the radius versus time of several tens of stars that were initially aligned on a radial vector, and subsequently orbited in accordance with equation (4). As time advances the shear will separate the particles in azimuth. In Figure 15 that factor has been removed, and so we see that in their radial motions the stars behave exactly like those in the circular wave of a colliding ring galaxy (see review of \citealt{ap96}).

\begin{figure}
\centerline{
\includegraphics[scale=0.27]{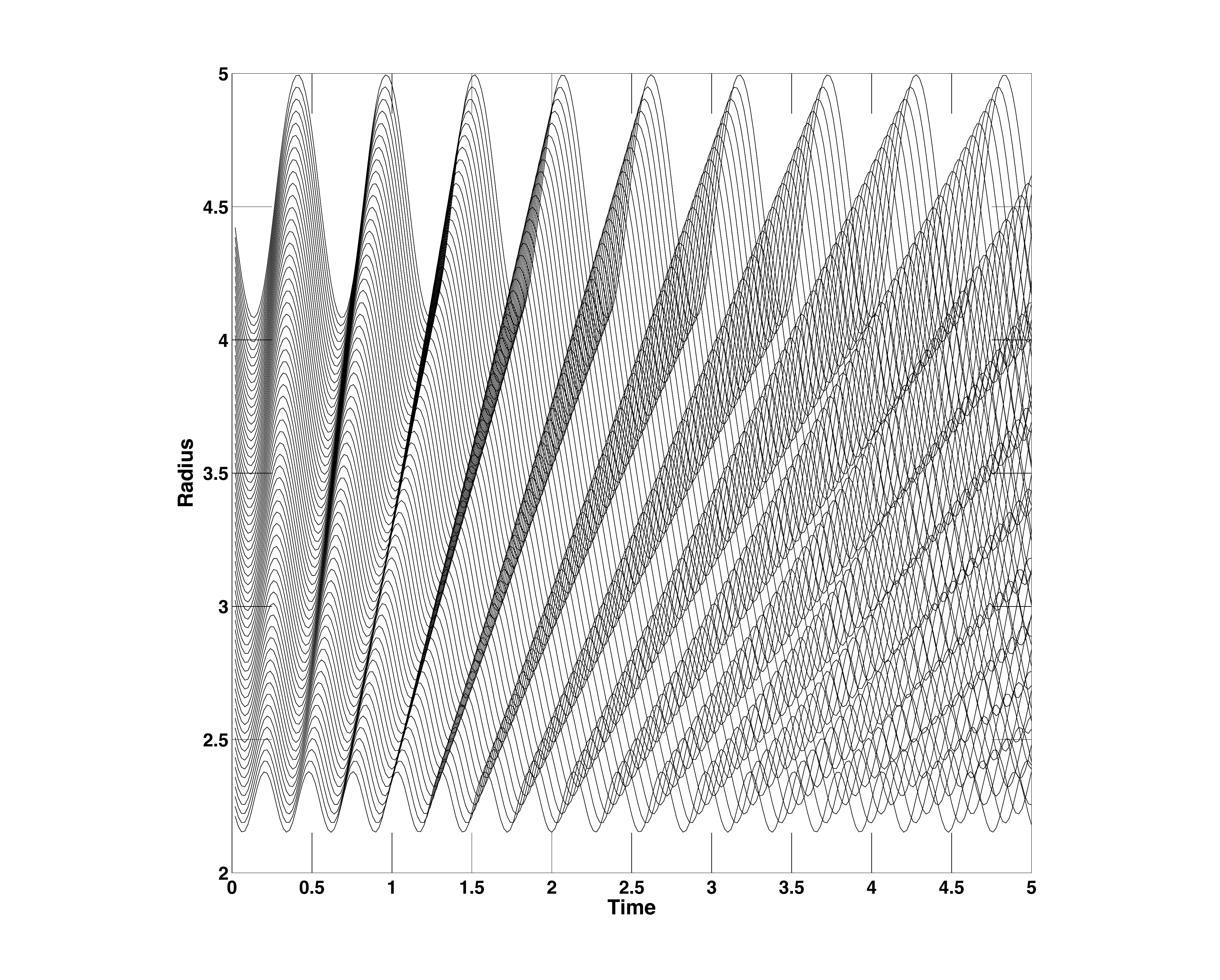}}
\caption{Radius versus time trajectories of sample particles from the analytic disc of Figure 13. All particles were on a single radial line at the time of the perturbation (t = 0). The subsequent azimuthal shear is not shown in this diagram, which illustrates how the epicyclic phase drift leads to the development of orbit-crossing zones with caustic boundaries as in ring galaxies.}
\end{figure} 

Like ring waves, Figure 15 shows that the epicyclic phase drift between orbits of different initial radii can be great enough to generate high-density orbit crossing zones. The earlier figures show that it takes a long time for shear to erase this behavior since it occurs over small ranges of radii. Figure 15 also explains the delayed build up of the peak wave densities. The phase drift takes some time to produce orbit crossing and caustics; initially the waves are only zones of moderate orbit compression. Eventually, the waves overlap and phase mix away in radius. Our experience with ring galaxies teaches us that strong perturbations form caustics much more quickly, and phase mix away equally quickly. However, with more moderate perturbations waves continue to be visible for a long time.

To the degree that the wave centre curves fit the analytic and numerical spirals (of Figs. 12 and 13), they provide an algebraic equation for those spirals and their evolution with time. The shape of these waves is derived, with approximations, from first principles. Later in their evolution these are tightly wound spirals, but they are not so tightly wound in the early stages, as evident in the figures.

We reemphasize that the waves described by the equations above are  kinematic waves, and the textbook results on kinematic waves apply to them (e.g., Sec. 6.1.3 of \citealt{bt}). Equation (7) is of the form usually adopted for kinematic spirals, with the last term being a specific form of the so-called shape function, derived from the caustic conditions. The pitch angle of kinematic spirals is generally found to be constant with radius, and that result should apply here too. It is well known that observed two-armed spirals can generally be approximately fit with constant pitch, logarithmic spirals, e.g., \citet{Elmegreen1989}.  Like \citet{Oh2008}, we find that the pitch angles also do not vary greatly with radius in any of our models. Pitch angles do vary with time, as discussed above for the models of Set 2. That is, the classical winding problem remains, though the caustic nature of the waves helps retain their strength and remain distinct for longer than might be expected.

\section{Conclusions}
We have shown that interactions between galaxies can produce grand
design spiral structure which persists for  at least many hundreds of Myr and many orbital periods (e.g., Fig. 1). In the self-consistent Set 2 simulations, with a
companion of 10 \% mass or more of the main galaxy, the waves
last for about a Gyr or more, at which point the companion may be a few
100 kpc away. These results are in accord with those of \citet{Oh2008}. We find that the mass of the companion needs to be at
least 1\% of the mass of the main galaxy to have a noticeable
effect. To produce an unbound orbit, we adopted a high velocity for
the perturber, which then slows after passing the main galaxy. Thus
the $m=2$ perturbation only fully develops a couple of 100 Myr or so
after closest approach. 

All of our models show that the spiral pattern slowly winds up over time,
thus at later times, it is difficult to see spiral structure in the
inner regions (in the stars at least) of the model discs. Thus, the longevity of our spirals is not the result of overcoming the classical windup problem. Rather it is because windup takes a long time outside the central disc, and because the contrast between wave crests and troughs is maintained for unexpectedly long times due to the caustic structure of the waves. In contrast, swing amplification can strongly enhance waves for a time, but in the long run the consequent restructuring destroys any correlations in the orbital parameters (e.g., epicyclic phase) responsible for the waves studied here. Swing amplification and caustic windup are two quite distinct evolutionary pathways for spiral waves. Some form of self-gravity has long been thought to be the most likely force to hold spiral waves in a quasi-steady state. However, the persistence of correlations resulting from tidal perturbations in kinematic waves with little self-gravity to destroy those correlations may work just as well, except perhaps in cases of recurrent swing amplification.

Tight spirals are reasonably well represented in catalogs of nearby galaxies, so they are evidently not too rare. NGC 488 is a prototypical example from the Hubble Atlas \citep{sa61}. The fact that these spirals take a number of typical rotation times to wind up (while maintaining a substantial density contrast) suggests that in most cases the companion will travel to large distances by the time the form becomes tightly wound. Thus, we do not expect most tightly wound spirals produced by this mechanism to have obvious close companions. In a group or cluster environment it may be hard to identify the collision partner. 

In some cases the collision partner may not be a galaxy. A disc galaxy falling past a cluster core at, for example, 2000 km s$^{-1}$ = (0.5 Mpc)/(250 Myr) may also experience a quasi-impulsive tidal perturbation from the cluster core as whole. This is in addition to fast galaxy-galaxy encounters likely to occur in the cluster. Similar considerations apply to galaxies falling into groups. The gravitational potential of the spiral waves generated by this mechanism will draw in gas locally, triggering star formation. This seems to be confirmed by the relatively modest-sized star forming knots observed in the waves of some nearby tightly-wound spirals. However, since the waves cover much of the disc, the integrated star formation could be significantly increased by these waves. 

We note that once the waves evolve to become tightly wound they may be very difficult to see directly, except in very nearby galaxies. The reader may inspect the various images of NGC 488 in the NASA Extragalactic Database. Many of them do not show the delicate wave structure visible in the HST image. We suspect that because this effect may be easy to induce in disc galaxies falling into larger structures, it plays a significant role in galaxy harassment and in causing the Butcher-Oemler effect. However, there is little direct evidence for it in observations of Butcher-Oemler clusters to date (\citealt{bu78}, but see \citealt{ya04} where some of the blue, ``passive'' cluster galaxies may be tight spirals), nor in the \citet{mo96} harassment simulations. In the former case it is very likely that the tightly wound waves could not be resolved. The numerical resolution may also be a factor in the latter case. Moreover, harassment involves multiple interactions, whose effects have not been studied in the context of long-lived waves.

We would expect that eventually higher order perturbations prevail in the
stellar disc or a bar forms. There is some indication of this at late times in the Set 2 models. In a lower resolution study,
we find the $m=2$ pattern does not last as long, indicating that the lifetime of the grand
design structure in numerical simulations will depend to some extent
on the resolution of the disc (see Appendix).

The mechanism described in this paper does not apply to strongly self-gravitating spirals, nor those driven by large-amplitude global disturbances, like bars. The classical theory of density waves shows that, when Q is low enough, a band of unstable wavelengths is present, where local self-gravity can overcome the effects of pressure and shear. Self-gravitating waves can then form (\citealt{bt}, \citealt{se10}). This would lead us to expect that weak triggering in a disc with a high value of Q would not be sufficient to produce strong waves. It would also predict the rapid dissipation of (kinematic) waves as they shear and wind up. The density enhancement in radial caustic zones overcomes these difficulties. The prediction that it operates in high-Q discs can be observationally checked. The mechanism is probably not applicable to young, gas-rich discs. Nonetheless, we reemphasize that it generates a class of nonlinear spirals whose structure and evolution are completely specified by a simple first-principles theory. For example, pattern speed and wind-up can be calculated straightforwardly given the structure of the potential (e.g., the unperturbed rotation curve). 

Because the inputs of this theory are so simple we also suspect that the waves may occur in other circumstances, and the theory may prove useful in applications beyond galaxy discs. Externally disturbed planetesimal accretion discs or Kuiper Belts might be possible areas of application. Even within the study of galaxy discs it should be a useful tool in addressing a variety of problems.

\section*{Acknowledgments}

We are grateful to Bev Smith, Jerry Sellwood, Alar Toomre, and the referee, H. M. Lee, for helpful comments. This research has made use of the NASA/IPAC Extragalactic Database (NED), which is operated by the Jet Propulsion Laboratory, Caltech, under contract with NASA. The calculations in Section 3 were performed on the University of Exeter's SGI Altix ICE 8200 supercomputer and the HLRB-II supercomputer at the Leibniz supercomputer centre. 

\bibliographystyle{mn2e}

\section{Appendix A: Lower Resolution Model}
We also ran a lower resolution calculation, using 90,000 gaseous disc particles, 50,000 stellar disc particles, 70,000  halo particles and 20,000 bulge particles, for the 0.3 ratio companion. We show the gas column density at times of 500 and 800 Myr in Figure~16. The $m=2$ perturbation is clear at 500 Myr. However by 800 Myr, the $m=2$ spiral structure is washed out by higher order perturbations (see Figure 16). For the higher resolution simulations, the $m=2$ structure persists for much longer, hence the lifetime of the two-armed spiral appears to be resolution dependent. 

\begin{figure}
\centerline{
\includegraphics[scale=0.25]{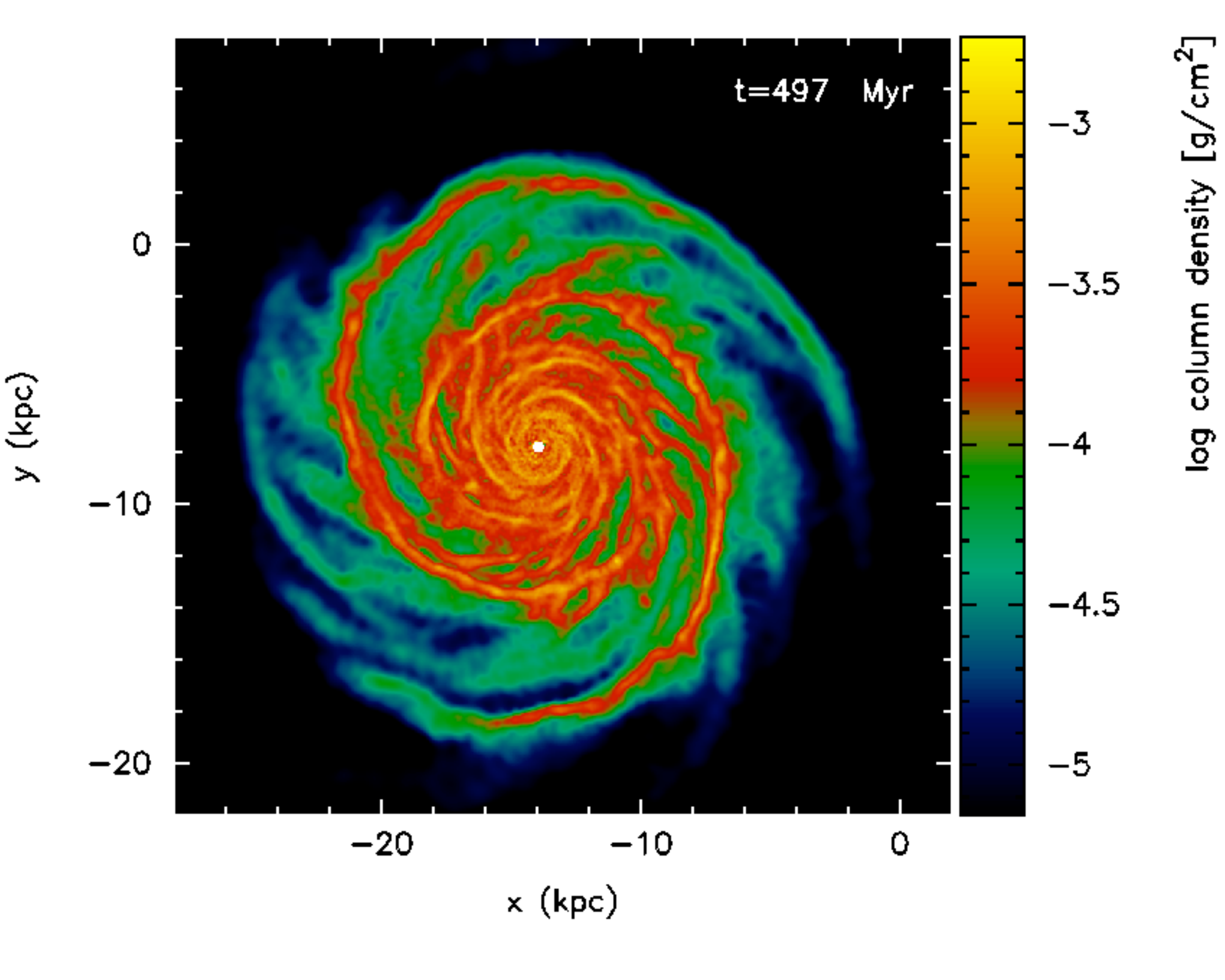}}
\centerline{
\includegraphics[scale=0.25]{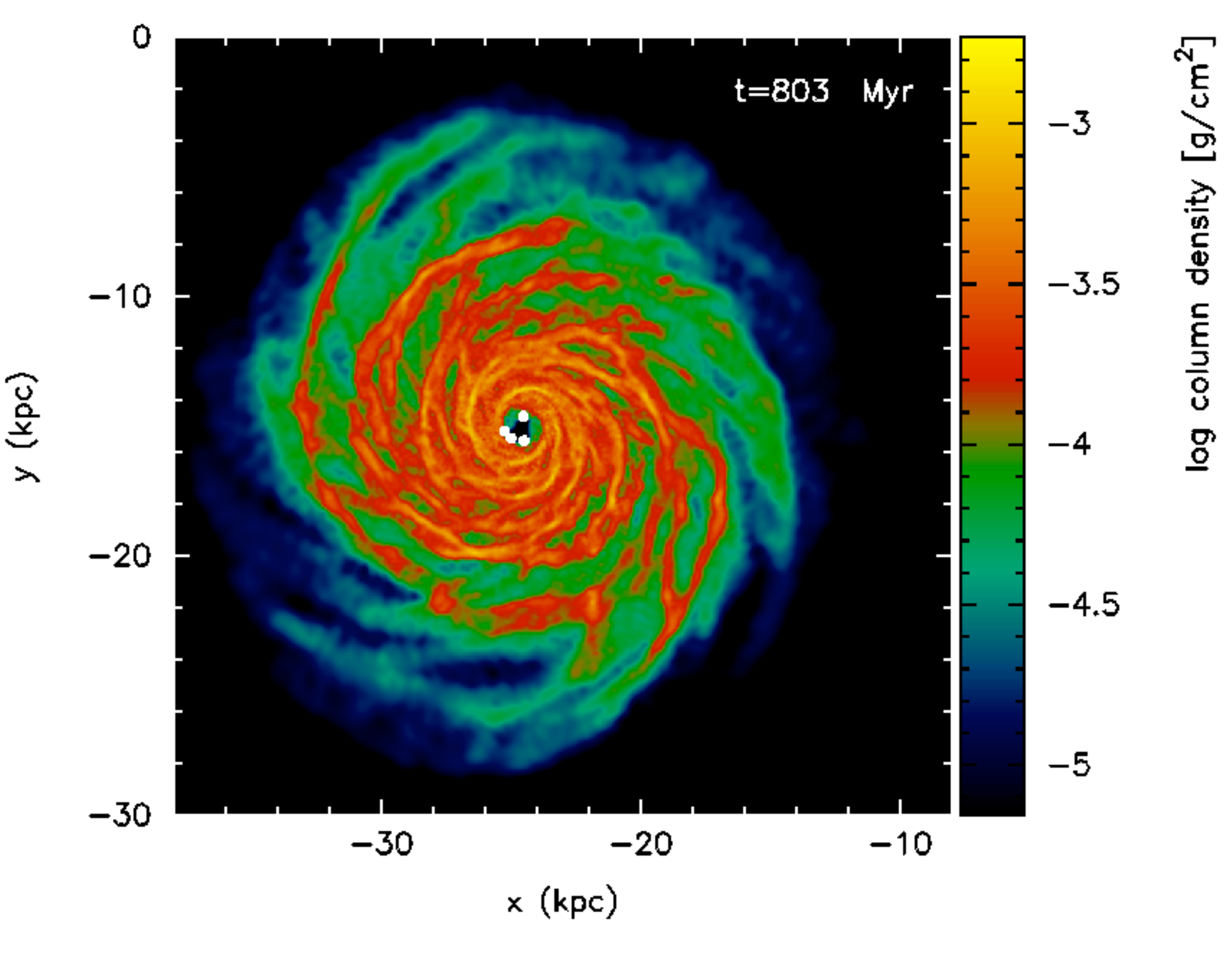}}
\caption{The gas column density is shown for a lower resolution simulation, with the 0.3 ratio companion at times of 500 and 800 Myr. The two-armed spiral structure present at 500 Myr is washed out by 800 Myr.}
\end{figure}

\bsp
\label{lastpage}
\end{document}